\documentclass[trackchanges,twocolumn]{aastex62}

\usepackage{amsmath}
\usepackage[caption=false]{subfig}

\received{}
\revised{}
\accepted{}
\submitjournal{ApJ}

\shorttitle{Supernova Photometric Classification}
\shortauthors{Villar et al.}

\begin{document}

\title{Supernova Photometric Classification Pipelines Trained on Spectroscopically Classified Supernovae from the Pan-STARRS1 Medium-Deep Survey}

\correspondingauthor{Ashley Villar}
\email{vvillar@cfa.harvard.edu}

\author[0000-0002-0786-7307]{V.~A.~Villar}
\affil{Center for Astrophysics \textbar{} Harvard \& Smithsonian, 60 Garden Street, Cambridge, MA 02138-1516, USA}

\author[0000-0002-9392-9681]{E.~Berger}
\affil{Center for Astrophysics \textbar{} Harvard \& Smithsonian, 60 Garden Street, Cambridge, MA 02138-1516, USA}

\author{G.~Miller}
\affil{Center for Astrophysics \textbar{} Harvard \& Smithsonian, 60 Garden Street, Cambridge, MA 02138-1516, USA}

\author{R.~Chornock}
\affil{Astrophysical Institute, Department of Physics and Astronomy, 251B Clippinger Lab, Ohio University, Athens, OH 45701-2942, USA}

\author{A.~Rest}
\affil{Space Telescope Science Institute, 3700 San Martin Dr., Baltimore, MD 21218, USA}

\author{D.~O.~Jones}
\affil{Department of Astronomy and Astrophysics, University of California, Santa Cruz, CA 92064,USA}

\author{M.~R.~Drout}
\affil{Department of Astronomy and Astrophysics, University of Toronto, 50 George Street, Toronto, Ontario, M5S 3H4 Canada}

\author{R.~J.~Foley}
\affil{Department of Astronomy and Astrophysics, University of California, Santa Cruz, CA 95064}

\author{R.~Kirshner}
\affil{Center for Astrophysics \textbar{} Harvard \& Smithsonian, 60 Garden Street, Cambridge, MA 02138-1516, USA}
\affil{Gordon and Betty Moore Foundation, 1661 Page Mill Road, Palo Alto, CA 94028}

\author{R.~Lunnan}
\affil{The Oskar Klein Centre \& Department of Astronomy, Stockholm University, AlbaNova, SE-106 91 Stockholm, Sweden}

\author{E.~Magnier}
\affil{Institute for Astronomy, University of Hawaii at Manoa, 2680 Woodlawn Dr., Honolulu, HI 96822, USA}

\author{D.~Milisavljevic}
\affil{Department of Physics and Astronomy, Purdue University, 525 Northwestern Avenue, West Lafayette, IN 47906, USA}

\author{N.~Sanders}
\affil{WarnerMedia Applied Analytics, 535 Boylston St., Boston, MA 02116}

\author{D.~Scolnic}
\affil{Department of Physics, Duke University, 120 Science Drive, Durham, NC, 27708, USA}

\begin{abstract}
Photometric classification of supernovae (SNe) is imperative as recent and upcoming optical time-domain surveys, such as the Large Synoptic Survey Telescope (LSST), overwhelm the available resources for spectrosopic follow-up. Here we develop a range of light curve classification pipelines, trained on 513 spectroscopically-classified SNe from the Pan-STARRS1 Medium-Deep Survey (PS1-MDS): $357$ Type Ia, $93$ Type II, $25$ Type IIn, $21$ Type Ibc, and $17$ Type I SLSNe. We present a new parametric analytical model that can accommodate a broad range of SN light curve morphologies, including those with a plateau, and fit this model to data in four PS1 filters ($g_\mathrm{P1}r_\mathrm{P1}i_\mathrm{P1}z_\mathrm{P1}$). We test a number of feature extraction methods, data augmentation strategies, and machine learning algorithms to predict the class of each SN. Our best pipelines result in $\approx 90\%$ average accuracy, $\approx 70\%$ average purity, and $\approx 80\%$ average completeness for all SN classes, with the highest success rates for Type Ia SNe and SLSNe and the lowest for Type Ibc SNe. Despite the greater complexity of our classification scheme, the purity of our Type Ia SN classification, $\approx 95$\%, is on par with methods developed specifically for Type Ia versus non-Type Ia binary classification.  As the first of its kind, this study serves as a guide to developing and training classification algorithms for a wide range of SN types with a purely empirical training set, particularly one that is similar in its characteristics to the expected LSST main survey strategy. Future work will implement this classification pipeline on $\approx 3000$ PS1/MDS light curves that lack spectroscopic classification.
\end{abstract}


\section{Introduction}

Optical time-domain astronomy has entered a new era of large photometric datasets thanks to current and upcoming deep and wide-field surveys, such as the Panoramic Survey Telescope and Rapid Response System (Pan-STARRS; \citealt{kaiser2010}), the Asteroid Terrestrial-impact Last Alert System (ATLAS; \citealt{jedicke2012atlas}), the Zwicky Transient Facility (ZTF; \citealt{kulkarni2018zwicky}), the Large Synoptic Survey Telescope (LSST; \citealt{ivezic2011large}), and the Wide Field Infrared Survey Telescope (WFIRST; \citealt{spergel2015}). The current surveys are already discovering $\sim 10^4$ SNe per year, a hundred-fold increase over the rate of discovery only a decade ago.  LSST will increase this discovery rate to $\sim 10^6$ SNe per year.

Supernovae have traditionally been classified based on their spectra \citep{filippenko1997optical}.  In the early days this was accomplished through visual inspection, then with template-matching techniques (e.g., SNID; \citealt{blondin2007determining}), and most recently with deep learning techniques (e.g., \citealt{muthukrishna2016deep}). However, given the current discovery rate, and the anticipated LSST discovery rate, spectroscopic follow up is severely limited.  The consequence of this fact is twofold.  First, we need a way to effectively identify ``needles'' in the haystack -- the events that will yield the greatest scientific return with detailed follow up observations (e.g., spectroscopy, radio, X-ray). Second, we need to devise methods to extract as much information and physical insight as possible from the ``haystack'' of SNe for which no spectroscopy or other data will be available. Here, we specifically focus on the latter issue and explore the question: Given complete optical light curves, can we classify SNe into their main spectroscopic classes (Ia, Ibc, IIP, etc.)?

\begin{figure*}[t!]
\begin{center}
\hspace*{0.1in}
{\includegraphics[width=\textwidth]{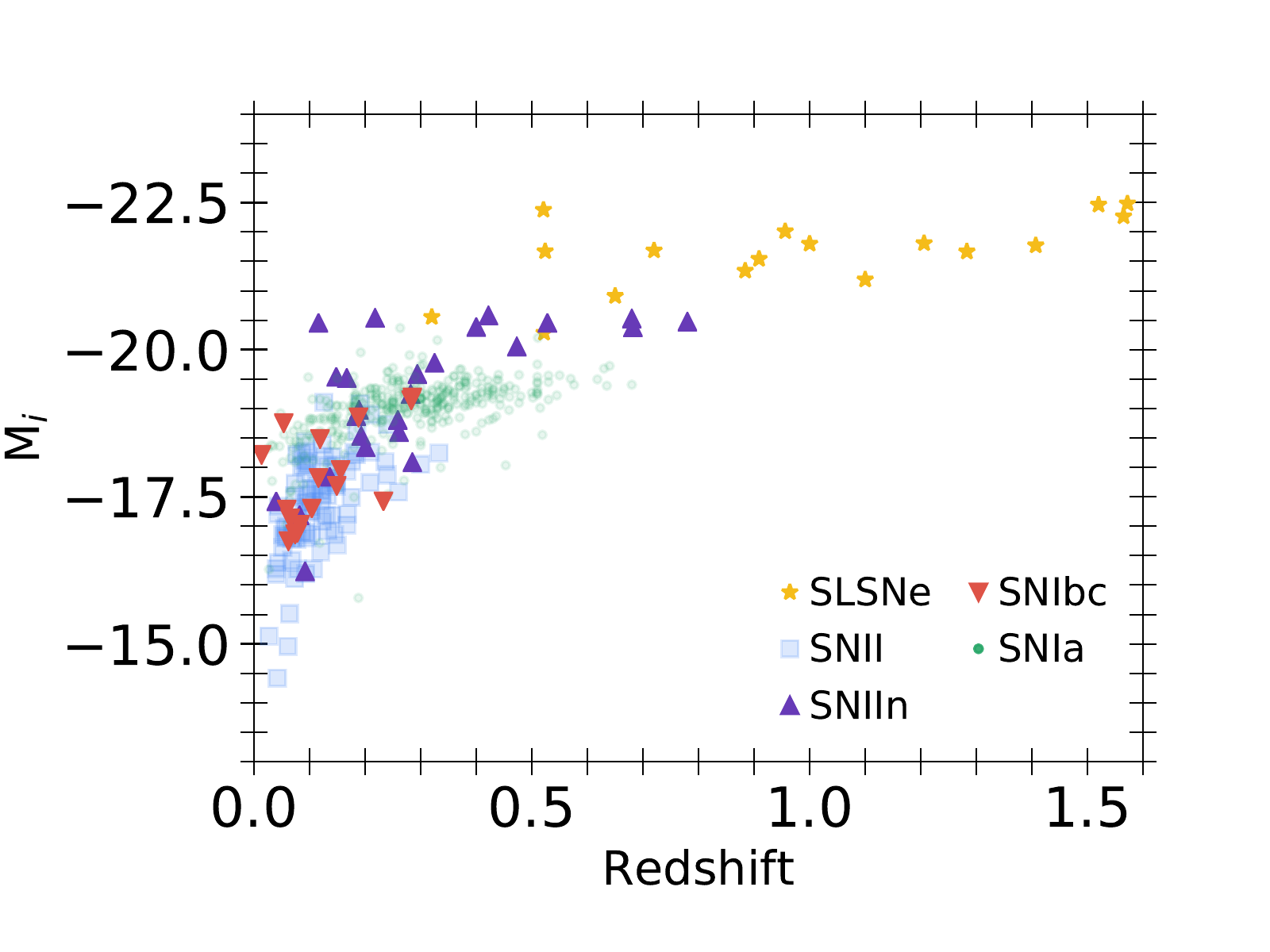}}
\vspace{-0.35in}
\caption{Peak $i_\mathrm{P1}$-band absolute magnitude versus redshift for the sample of PS1-MDS spectroscopically-classified SNe used in this study. We apply a cosmological $k$-correction to the peak magnitudes, but do not correct for the intrinsic spectral energy distribution of the various SNe. The sample includes five SN classes: Ia (green circle), Ibc (red downward triangle), II (blue sqaure), IIn (purple upward triangle), and SLSNe (yellow star).}
\end{center}
\label{fig:redmag}
\end{figure*}

Previous studies in this area have largely focused on the simpler task of separating thermonuclear Type Ia SNe from non-Type Ia SNe, motivated by the use of Type Ia SNe as standardizable cosmological candles, and taking advantage of their uniformity (e.g., \citealt{moller2016,kimura2017single}). Separating the classes of core-collapse SNe (CCSNe) is a broader and more challenging problem. First, unlike Type Ia SNe, CCSNe exhibit broad diversity between and within each class in terms of basic properties such as luminosity, timescale, and color (e.g.,  \citealt{drout2011first,taddia2013carnegie,sanders2015toward,nicholl2017magnetar,villar2017theoretical}).  This is due to their wide variety of progenitor systems, energy sources, and circumstellar environments. Second, the overall diversity of CCSNe is less thoroughly explored, due to small sample sizes and few published uniform studies.  As a consequence, most previous works on photometric classification of CCSNe have relied on simulated datasets to train and test classification algorithms (e.g., \citealt{richards2011semi,charnock2017deep,kimura2017single,ishida2018optimizing}). Simulated datasets are based on strong assumptions about the underlying populations of each SN class and often do not reflect the true event diversity, or the effects of actual survey conditions. 

Here, we approach the question of SN photometric classification using a large and uniform dataset of $513$ spectroscopically-classified SNe from the PS1-MDS.  Importantly, the characteristics of this dataset in terms of filters, depth, and cadence are the closest available analogue to the LSST main survey design. We fit the observed light curves with a flexible analytical model that can accommodate all existing light curve shapes, using a Markov chain Monte Carlo (MCMC) approach.  We then train and evaluate 24 classification pipelines that span different feature extraction, data augmentation, and classifications methods.  We further use the posteriors of our MCMC fits to determine overall uncertainties on our classifications. 

The paper is organized as follows. In \S\ref{sec:data} we introduce the PS1-MDS dataset utilized here. In \S\ref{sec:model} we describe our analytical light curve model and iterative MCMC fitting approach. In \S\ref{sec:classification} we describe the key components of our various classification pipelines, including feature extraction, data augmentation, and classification approaches. We present the results of our classifications in \S\ref{sec:results}, compare to previous classifications efforts in \S\ref{sec:comp}, and discuss limitations and future directions in \S\ref{sec:future}.

Throughout this paper, we assume a flat $\Lambda$CDM cosmology with $\Omega_M=0.286$, $\Omega_\Lambda=0.712$ and $H_0 = 69.3$ km s$^{-1}$ Mpc$^{-1}$ \citep{hinshaw2013nine}.

\section{PS1-MDS Supernova Light Curves and Spectroscopic Classifications}
\label{sec:data}

Pan-STARRS1 (PS1) is a wide-field survey telescope with a 1.8 m diameter primary mirror located on Haleakala, Hawaii \citep{kaiser2010}. The PS1 1.4 gigapixel camera (GPC1) is an array of 60 $4800\times 4800$ pixel detectors with a pixel scale of $0.''258$ and an overall field of view of $7.1$ deg$^2$. The PS1 survey used five broadband filters, $g_\mathrm{P1}r_\mathrm{P1}i_\mathrm{P1}z_\mathrm{P1}y_\mathrm{P1}$. The details of the filters and the photometry system are given in \citet{stubbs2010} and \citet{tonry2012pan}.

The PS1-MDS, conducted in $2010-2014$, consisted of ten single-pointing fields for a total area of about $70$ deg$^2$ \citep{chambers2016}. About $25\%$ of the overall survey observing time was dedicated to the MDS fields, which were observed with a cadence of about 3 days per filter in $g_\mathrm{P1}r_\mathrm{P1}i_\mathrm{P1}z_\mathrm{P1}$ to a $5\sigma$ depth of $\approx 23.3$ mag per visit. The typical sequence consisted of $g_\mathrm{P1}$ and $r_\mathrm{P1}$ on the same night, followed by $i_\mathrm{P1}$ and then z$_\mathrm{P1}$ on subsequent nights. Observations in $y_\mathrm{P1}$-band were concentrated near full moon with a shallower $5\sigma$ depth of $\approx 21.7$ mag; we do not use the $y_\mathrm{P1}$-band data in this study due to its significantly shallower depth and poorer cadence. 

The reduction, astrometry, and stacking of the nightly images were carried out by the Pan-STARRS1 Image Processing Pipeline (IPP; \citealt{magnier2016a,magnier2016b,waters2016pan}). The nightly stacks were then transferred to the Harvard FAS Research Computing Odyssey cluster for a transient search using the {\tt photpipe} pipeline, previously used in the SuperMACHO and ESSENCE surveys \citep{rest2005testing,miknaitis2007essence} and described in detail in our previous analyses of PS1-MDS data \citep{rest2014,scolnic2017complete,jones2018measuring}. 

In the full PS1-MDS dataset we have identified 5235 likely SNe \citep{jones2017measuring,jones2018measuring}.  During the course of the survey, spectroscopic observations were obtained for over 500 events using the MMT 6.5-m telescope, the Magellan 6.5-m telescopes, and the Gemini 8-m telescopes.  We further obtained spectroscopic host galaxy redshifts for 3147 SN-like transients. The transients spectroscopically and photometrically classified as Type Ia SNe were published in \citet{jones2017measuring}; the light curves and photometric classification of the remaining objects will be presented in future work. Similarly, the bulk of the Type IIP SNe (76 events) were published in \citet{sanders2015toward}, and the Type I SLSNe (17 events) were published in \citet{lunnan2018}. Here we focus on $513$ spectroscopically classified events, which were classified using the SNID software package \citep{blondin2007determining}.  The sample contains $357$ Type Ia SNe, $93$ IIP/L SNe, $25$ Type IIn SNe, $21$ Type Ibc SNe, and $17$ Type I SLSNe\footnote{Three of the 17 SLSNe (PS1-12cil, PS1-10ahf, and PS1-13or) do not have spectroscopic host redshift measurements. \citet{lunnan2018} estimated their redshifts (0.32, 1.10 and 1.52, respectively) from strong rest-frame UV features for the $z>1$ objects and SN Ic-like post-peak features for PS1-12cil.}. 

Our sample is limited events with high-confidence spectroscopic classifications with a statistically useful number of members in each class. As part of the PS1-MDS we discovered several other rare transients, including tidal disruption events \citep{gezari2012ultraviolet,Chornock2014} and fast-evolving luminous transients \citep{drout2014rapidly}, but the sample sizes for those are too small for inclusion in this study. It is possible that SNID misclassification exist in our dataset; e.g., low SNR events are more likely to match to Type Ia SNe \citep{blondin2007determining}. To partially counteract this, we check each member of our Type Ibc and SLSNe classes (our smallest classes) by eye to ensure high purities. Finally, we note that the magnitude limit for our spectroscopic follow up was generally shallower by about 1.5 mag relative to the PS1-MDS nominal per-visit depth. This does not affect our ability to test classifiers on the spectroscopic sample itself, but  will be considered when extending our method to the full photometric dataset in future work (see \S\ref{sec:comp}).

The light curves range from a minimum of 3 to $\approx 150$ total data points in any filter with a signal-to-noise ratio of ${\rm S/N}>3$, with a median of about 30 data points in each light curve. We have only eliminated events with light curves that contain fewer than two $3\sigma$ detections in three or more filters, eliminating $7$ SNe from our sample\footnote{For completeness, we ran our final classifier on these light curves as well and found that 5 of the 6 Type Ia SNe, as well as the one Type II SN were actually correctly classified, albeit with a low classification confidence.} (6 Type Ia SNe and 1 Type II SN) leaving 506 remaining SNe for our training set.

In Figure~\ref{fig:redmag} we plot the peak absolute $i_\mathrm{P1}$ magnitude versus redshift for our spectroscopic sample. The sample spans $M_i\approx -14.5$ to $-22.5$ and extends in redshift to $z\approx 1.6$, with only the brightest classes (SLSNe and Type IIn) being observed at $z\gtrsim 0.6$.  Specifically, we find a range of $M_i\approx -14.5$ to $-18.5$ mag for the Type II SNe, $\approx -16.5$ to $-19.5$ mag for the Type Ibc SNe, $-16$ to $-20.5$ for the Type IIn SNe, and $\approx -20.5$ to $-22.5$ for the SLSNe.

\section{Analytical Light Curve Model and Fitting}
\label{sec:model}

\begin{figure}[t!]
\begin{center}
\vspace{-0.1in}
\hspace*{-0.2in}
\scalebox{1.}
{\includegraphics[width=0.53\textwidth]{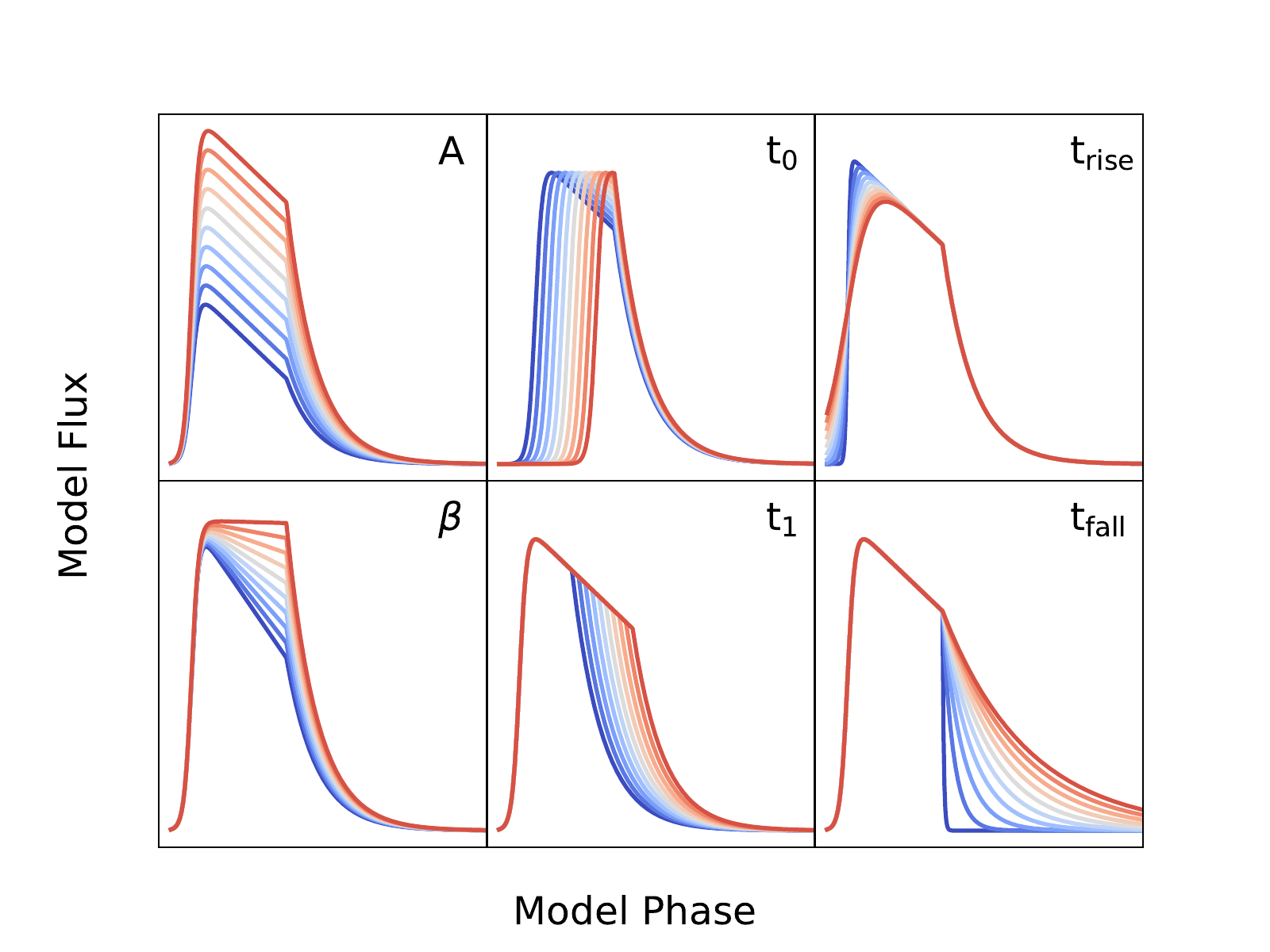}}
\caption{Example model light curves based on Equation~\ref{eqn:model} highlighting how each of the free parameters affects the light curves. The parameters are individually varied from low (blue) to high (red) values.}
\label{fig:lc_example}
\end{center}
\end{figure}

\begin{figure*}[t!]
\begin{center}
{\includegraphics[width=\textwidth]{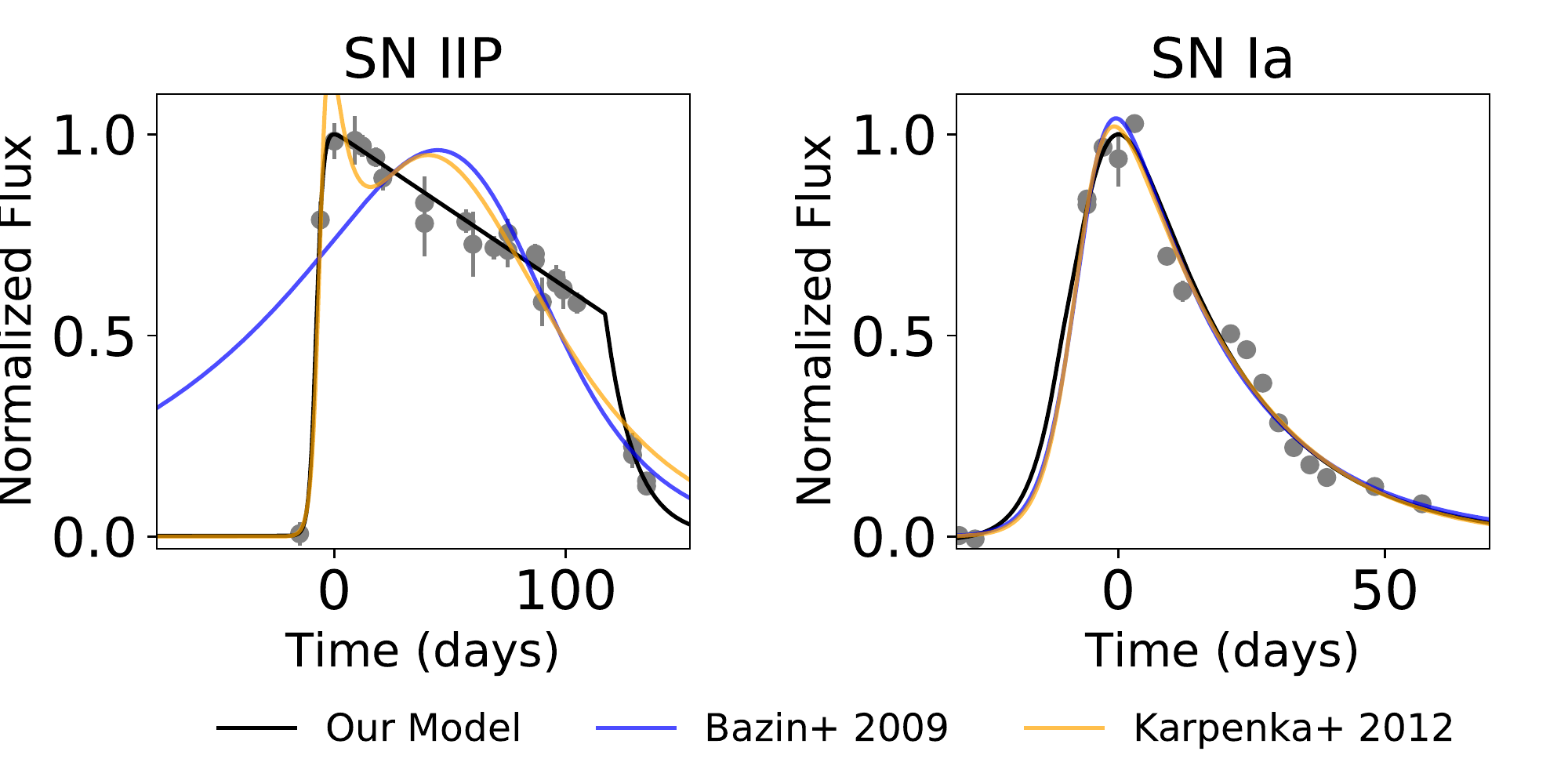}}
\caption{A comparison of our analytical light curve model (Equation~\ref{eqn:model}; black line) to that of \citet{bazin2009core} (blue line) and \citet{karpenka2013} (yellow line) for i-band lightcurves of both a Type IIP SN ({\it Left}) and a Type Ia SN ({\it Right}). Our model performs similarly for a Type Ia SN, but is superior at fitting SNe with a light curve plateau.}
\label{fig:lc_iip}
\end{center}
\end{figure*}

Rather than interpolating data points, a common method to standardize data is to fit a simple parametric model to the light curves (e.g., \citealt{bazin2009core,newling2011statistical,karpenka2013}). However, the majority of existing analytical light curve models are best-suited for Type Ia SNe and have limited flexibility for the full observed range of SN light curve shapes.  Here we present and fit our data with a new parametric piecewise model that is designed to be flexible enough for a broad range of light curve morphologies:

\begin{equation}
\label{eqn:model}
F=\begin{cases} 
      \frac{A+\beta(t-t_0)}{1+e^{-(t-t_0)/\tau_\mathrm{rise}}} & t<t_1 \\
      \frac{\left(A+\beta\left(t_1-t_0\right)\right)e^{-(t-t_1)/\tau_\mathrm{fall}}}{1+e^{-(t-t_0)/\tau_\mathrm{rise}}} & t\ge t_1 
   \end{cases}
\end{equation}

The model contains seven free parameters, whose effects on the resulting light curves are shown in Figure~\ref{fig:lc_example}. Although each parameter has a unique and interpretable effect, some degeneracies between the parameters exist. For example, the parameter $A$ affects the amplitude of the light curve, although its value does not exactly correspond to the peak flux. Similarly, $t_0$ acts as a temporal shift in the light curve, but does not directly correspond to the time of explosion or the time of peak. The parameters, $t_\mathrm{\rm rise}$, $t_1$, and $t_\mathrm{\rm fall}$ control the rise, plateau onset, and fall time of the light curve, respectively. For the purposes of fitting, we reparameterize $t_1$ into a new parameter $\gamma\equiv t_1-t_0$, which better represents the plateau duration of the light curve and results in fewer degeneracies when fitting. Finally, the parameter $\beta$ controls the slope of the plateau phase. 

This functional form is similar to those presented in \citet{bazin2009core} (with five free parameters) and \citet{karpenka2013} (with six free parameters), but incorporates a plateau component.  In Figure~\ref{fig:lc_iip} we show examples of fits to a Type IIP SN and a Type Ia SN  with our model, the Bazin model and the Karpenka model.  Our model provides a better fit to both the fast rise time and plateau phase of the Type IIP SN light curve, and is flexible enough to also fit the smoother light curve of a Type Ia SN.  We note that \citet{sanders2015toward} presented a similar piecewise model with 11 free parameters to fit a sample of 76 PS1-MDS Type II SNe; however, \citet{lochner2016photometric} found that this model was not robust when fitting data without the use of informative priors, due to the large number of free parameters. Additionally, the sharp transitions between rise and decline in the Sanders model make it difficult to fit CCSNe with smooth peaks.

\begin{deluxetable}{ccc}[t!]
\tabletypesize{\footnotesize} 
\tablecolumns{3} 
\tablewidth{0pt} 
\tablecaption{Parameter Descriptions and Priors \label{table:results1}} 
\tablehead{ 
\colhead{Parameter} & \colhead{Description} & \colhead{Prior}
} 
\startdata 
$\tau_\mathrm{rise}$ (days) & Rise Time & U(0.01,50)\\
$\tau_\mathrm{fall}$ (days) & Decline Time & U(1,300)\\
$t_0$ (MJD) & ``Start'' Time & U($t_\mathrm{min}-50$, $t_\mathrm{max}+300$\\
$A$ & Amplitude & U(3$\sigma$,100 $F_\mathrm{max}$)\\
$\beta$ (flux/day) & Plateau slope & U($-F_\mathrm{max}/150$,0)\\
$c$ (flux) & Baseline Flux & U($-3\sigma$,$3\sigma$)\\
$\gamma$ (days) & Plateau duration & $(2/3)N(5,5)+(1/3)N(60,30)$
\enddata 
\vspace{-0.8cm} 
\end{deluxetable}

One common Type Ia SN light curve feature missing from our model is the second peak in the red light curves at about 1 month post-explosion (e.g., \citealt{kasen2006secondary,mandel2011type,dhawan2015near}). We find that this feature manifests itself as a ``plateau'' in our analytical model in the $i$- and $z$-bands. However, as we show in \S\ref{sec:results}, our classification pipelines can reliably classify Type Ia SNe without explicitly including a second peak in our model.

We fit the light curves using PyMC (V2; \citealt{patil2010pymc}), a Python module that implements a Metropolis-Hastings MCMC sampling algorithm. We assume uniform priors on all parameters with the exception of $\gamma$. We found that light curves typically fall in one of two solutions: light curves with a long plateau (in the case of Type IIP SNe) and light curves that lack a plateau (all other types). To best reflect this fact, we set the prior of $\gamma$ to a double Gaussian peaked at 5 and 60 days. This prior helps to remove a degeneracy in which a steep exponential decline can resemble a linear decline. The priors are listed in Table~\ref{table:results1}. We use a standard likelihood function, incorporating both the observational error and a scalar white noise scatter term added in quadrature.

We find that several of our model parameters are correlated (degenerate) with one another. In particular, the amplitude ($A$) is negatively correlated with both the rise time ($t_\mathrm{rise}$) and plateau duration ($\gamma$) but negatively correlated with the start time ($t_\mathrm{0}$). Additionally, duration is negatively correlated to both the rise time and start time, while the rise time is positively correlated with the start time.

\begin{figure}[t!]
\begin{center}
\hspace*{-0.125in}
\scalebox{1.}
{\includegraphics[width=0.49\textwidth]{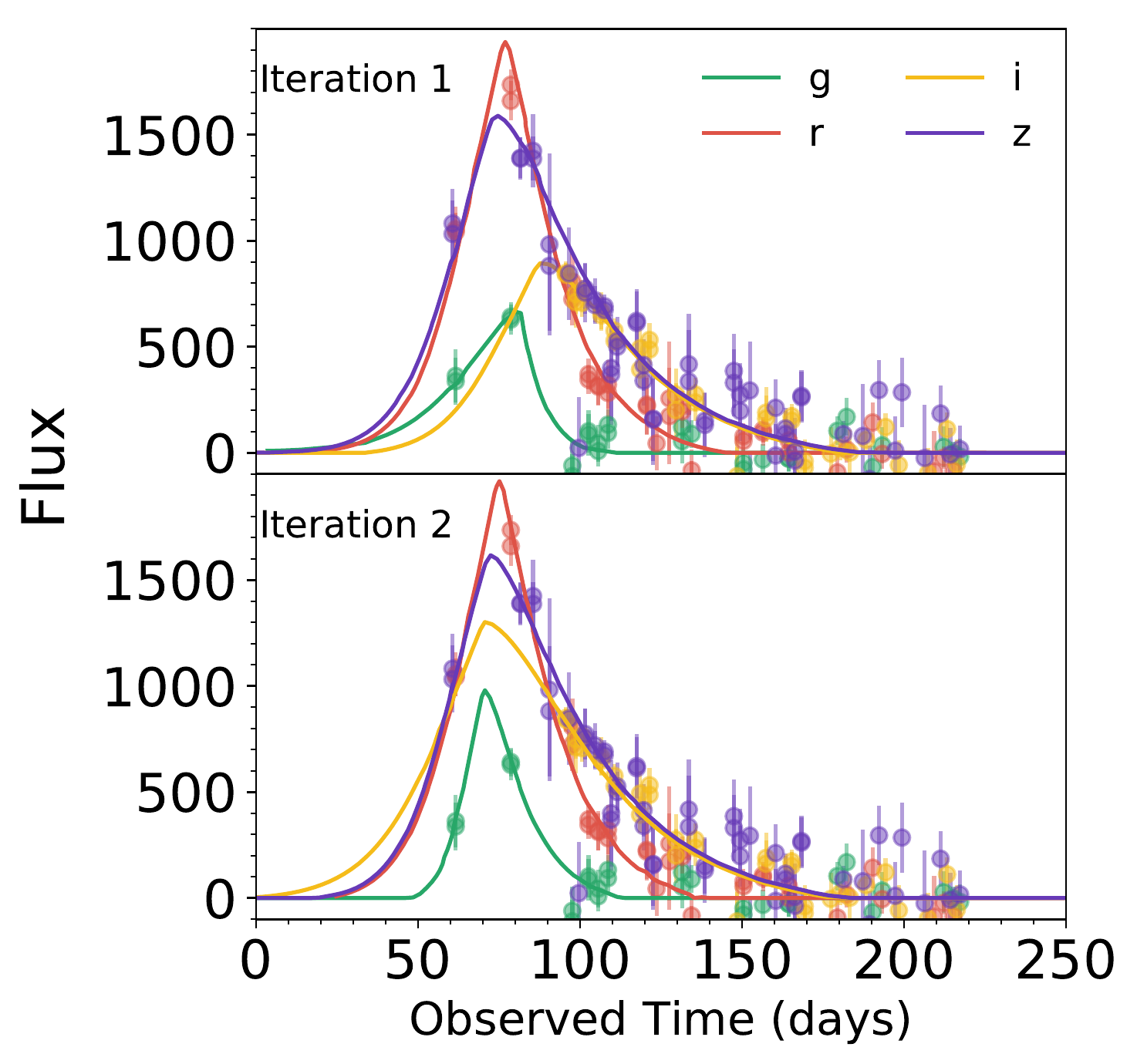}}
\caption{Example best-fit light curves in the 4 PS1 filters after the first (top) and second (bottom) MCMC iterations. Following the first iteration, the peak time varies significantly between the filters due to differences in the data quality and time sampling.  The best-fit solution of the second iteration, using the combined posteriors from the first iteration, provides much better agreement in the light curve properties.}
\label{fig:lc_its}
\end{center}
\end{figure}

We fit the light curve in each of the 4 filters independently, in the observer frame, but use an iterative fitting routine to incorporate combined information from all filters. We first run the MCMC to convergence on each filter independently with the same set of priors. We then combine the marginalized posteriors (i.e., we ignore parameter covariances) from each filter and use the combined posterior as a new prior for a second iteration of fitting.  We can apply this process repeatedly, but we find that a single iteration is sufficient for the vast majority of events.  Our iterative procedure is essential for fitting light curves in which some filters have significantly fewer data points, a situation that is common in photometric surveys due to differences in relative sensitivity, the intrinsic colors and color evolution of SNe, and varying observing conditions. An example of the best-fit solutions given by the first and second iterations is shown in Figure~\ref{fig:lc_its}. In this example, the peak times in $g$- and $i$ are in disagreement with $r$- and $z$ due to poorly-sampled data in the former two filters. Following the second iteration, this disagreement is removed, leading to more realistic fits.

\begin{figure*}[t!]
\begin{center}
\hspace*{-0.2in}
\scalebox{1.}
{\includegraphics[width=1.1\textwidth]{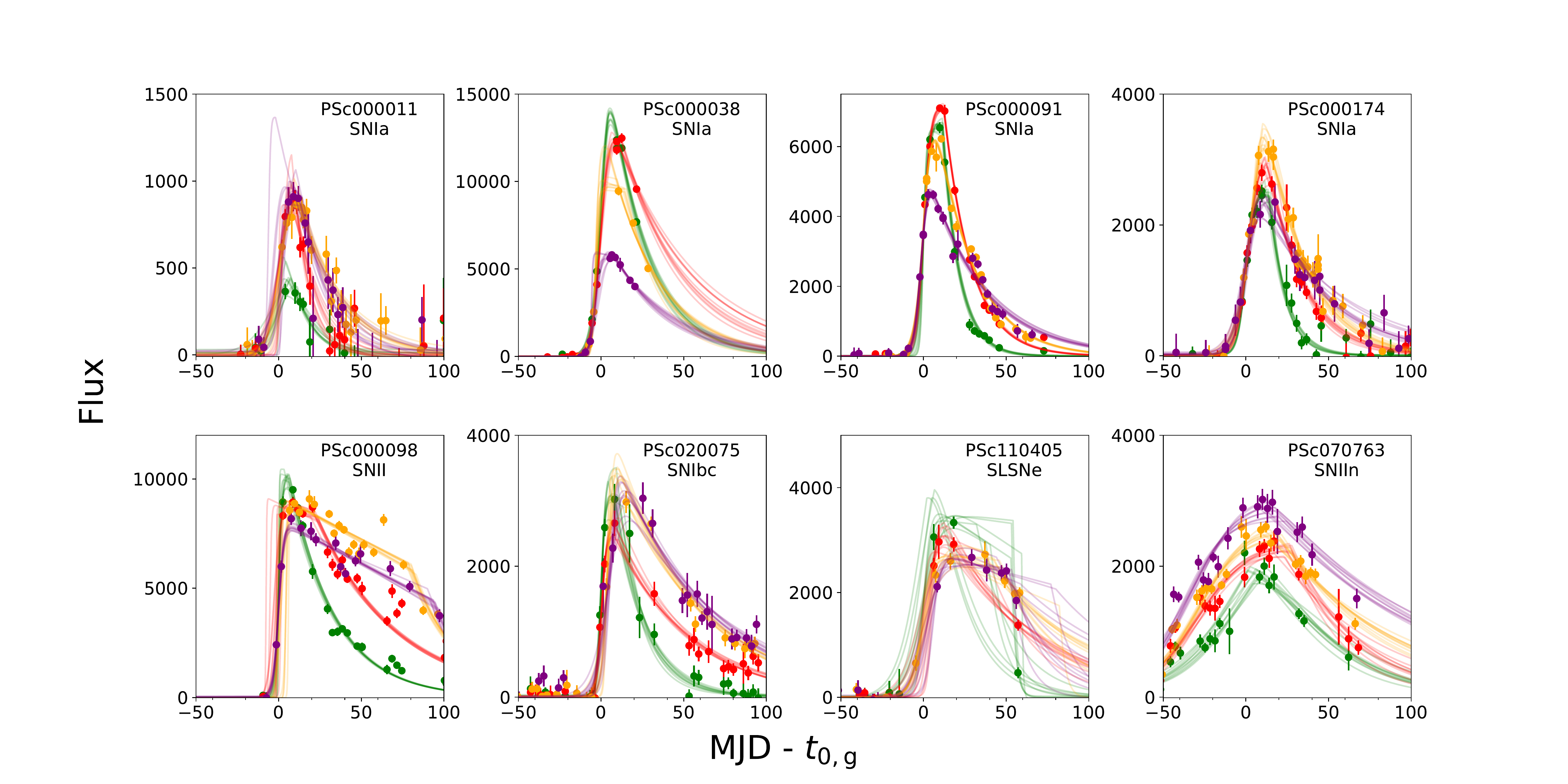}}
\caption{Example light curves and sample posterior draws of associated model fits in the 4 filters for various SN types.  The model is described by Equation~\ref{eqn:model} and the fitting procedure is described in \S\ref{sec:model}.}
\label{fig:lc_examples}
\end{center}
\end{figure*}

Representative light curves and their best fits are shown in Figure~\ref{fig:lc_examples}. The solutions are constrained for well-sampled light curves (e.g., the Type Ia SN shown) but more poorly constrained for sparse light curves (e.g., the SLSN shown). Crucially, because we have access to the full posterior of light curve solutions, we can feed many samples of the posterior through our classification algorithm to quantify the classification uncertainty for each event.

Unless otherwise specified, we use the observer-frame light curve fits to extract features. We then include the redshift to transform to absolute magnitudes, including a cosmological $k$-correction: $M=m-5\log(d_L/10\mathrm{pc})+2.5\log(1+z)$, where $d_L$ is the luminosity distance.  We do not apply $k$-corrections to account for the intrinsic spectral energy distribution of the various SN types.

\section{Classification Pipelines}
\label{sec:classification}

For each SN, our MCMC fitting generates posterior distributions for the model light curve parameters. To train a classifier, we need to extract features from the light curves generated by the fitted parameters. We test several methods of feature extraction, data augmentation and classification. We describe each method in the following subsections, and we compare the algorithms in terms of classification purity, completeness, and accuracy in \S\ref{sec:results}. \textit{Purity} (also called \textit{precision}) is defined as the fraction of events in a predicted class that are correctly identified; for example, if our classifier predicts a total of 100 Type Ia SNe, but only 70 of those are spectroscopically-classified as Type Ia SNe, the purity would be 0.7. \textit{Completeness} (also called \textit{recall}) is defined as the fraction of events in an observed class that are correctly identified; for example, if our sample contains 100 spectrosopically-classified Type Ia SNe, but our classifier has only identified 70 of those events as Type Ia SNe, then our completeness would be 0.7.  \textit{Accuracy} is defined as the total fraction of events that are classified correctly as being a member or not a member of a given class; for example, if a sample of 100 SNe contains 70 spectrosopically-classified Type Ia SNe, and our classifier correctly identifies the 70 Type Ia SNe but incorrectly classifies 20 more CCSNe as Type Ia SNe, the overall accuracy is 0.8. The three terms are mathematically defined as follows:
\begin{align}
  \mathrm{Purity}            &= \frac{\mathrm{TP}}{\mathrm{TP+FP}}\\ \mathrm{Completeness}      &= \frac{\mathrm{TP}}{\mathrm{TP+FN}}\\
  \mathrm{Accuracy}          &= \frac{\mathrm{TP+TN}}{\mathrm{TS}},
\end{align}
where TP (FP) is the number of true (false) positives, TN (FN) is the number true (false) negatives, and TS is the total sample size.

\subsection{Feature Selection}
\label{sec:features}

Although our analytical model produces interpretable features for each light curve (albeit ones that are somewhat degenerate) we would like to explore various methods of feature extraction, based on the analytical fits.  In particular, we explore the following four types of features:

\begin{itemize}
    \item \textbf{Model Parameters (M):} We use the analytical model parameters as features, as well as the peak absolute magnitude in each filter, including a cosmological $k$-correction but no correction for intrinsic SN colors and color evolution. 
    \item \textbf{Hand-Selected Features (HS):} We use hand-selected interpretable features: the peak absolute magnitude in each filter, including a cosmological $k$-correction but no correction for intrinsic SN colors and color evolution; and the rest-frame rise and fall times by 1, 2 and 3 mag relative to peak (where we do \textit{not} correct the rise and fall times for cosmological time-dilation). 
    \item \textbf{Principal Component Analysis (PCA):} We fit a PCA decomposition model to the full set of analytical model fits (without any redshift corrections) independently for each filter.  We use the first 6 PCA components from each filter, corresponding to an explained variance within the light curves of $\sim 99.9$\%.  We also use the peak absolute magnitude, including a cosmological $k$-correction, in each filter in addition to the PCA components. 
    \item \textbf{Light Curves (LC):} We use the model light curves as the features. We renormalize the flux of each light curve, correcting for luminosity distance; however, we find that neglecting time dilation corrections improves classification accuracy, and therefore we do not make these corrections. We down-sample each filter model to 10 observations logarithmic-spacing between $t_0$ and $t_0+300$ to decrease the number of features.
\end{itemize}

\begin{figure*}[t]
\begin{center}
\hspace*{-0.1in}
\scalebox{1.}
{\includegraphics[width=\textwidth]{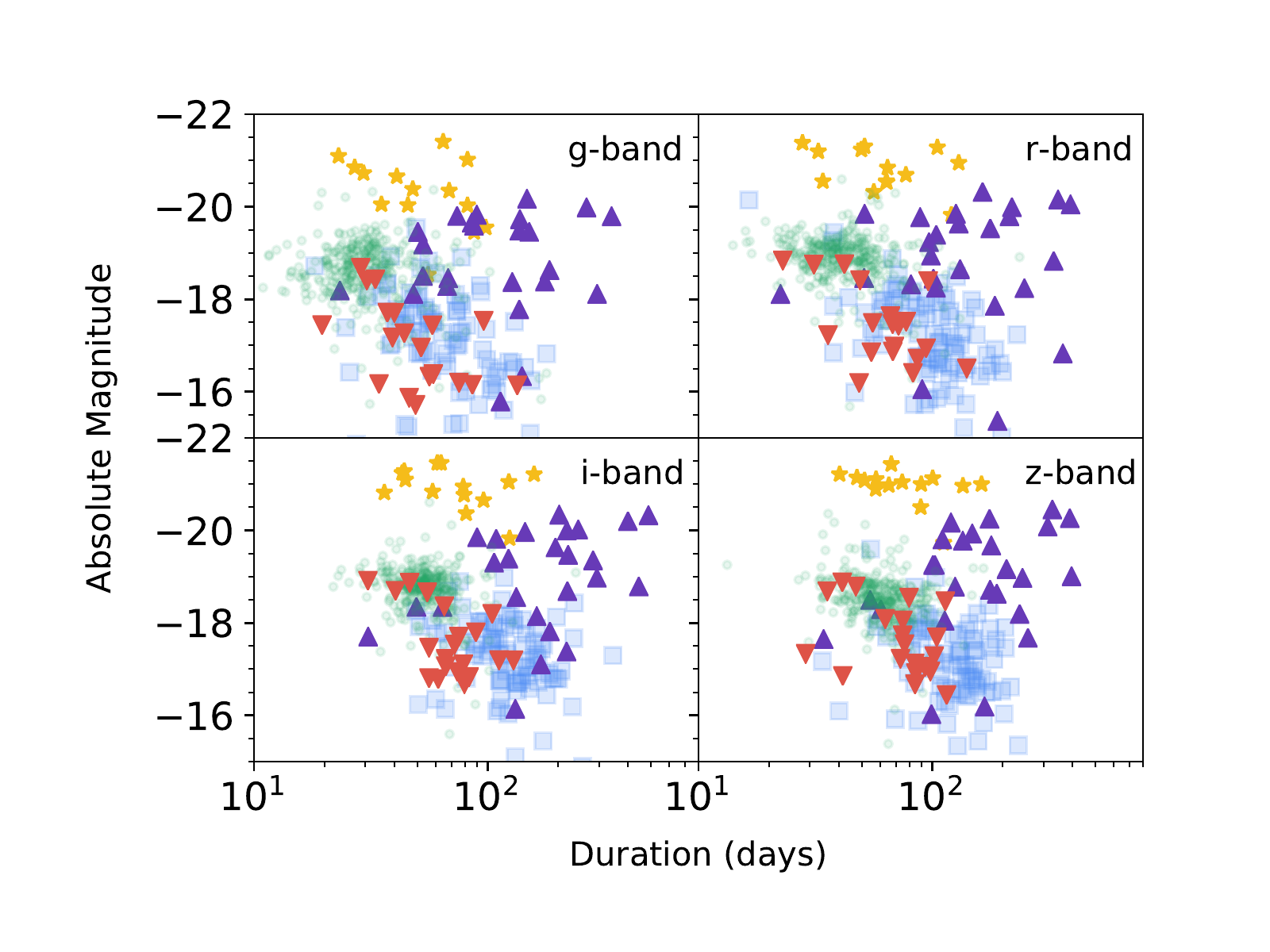}}
\vspace{-0.3in}
\caption{Duration-luminosity feature space of the dataset in the 4 PS1 filters. Duration is defined as the total time for the light curves to rise and decline by 2 mag relative to the peak.  The plotted values are from the median model fits to the light curves using Equation~\ref{eqn:model}. The sample includes the five SN classes: Ia (green circle), Ibc (red downward triangle), II (blue square), IIn (purple upward triangle), and SLSNe (yellow star).}
\label{fig:dlps}
\end{center}
\end{figure*}

To provide some intuition, we highlight a sub-space of the hand-selected features ($M_{\rm peak}$ versus duration time to rise and fall by 2 mag) in Figure~\ref{fig:dlps}.  We find that some SN classes, such as SLSNe versus Type II, or Type Ia versus Type IIn, easily separate in the duration-luminosity feature space. However, other classes, such as Type Ibc versus Type Ia and IIP, have substantial overlap in this space, regardless of filter. This highlights that while simple heuristics can be used as first-order classifiers for some SN classes, other classes are intrinsically difficult to disentangle from light curve information alone.

\subsection{Data Augmentation}
\label{sec:augmentation}

Data augmentation is ubiquitous in machine learning applications, as a larger dataset can significantly improve the accuracy and generalizability of most classification algorithms. Data augmentation methods have already been utilized in the astrophysical context (e.g., \citealt{hoyle2015data}). 

\begin{figure*}[t!]
\begin{center}
{\includegraphics[width=\textwidth]{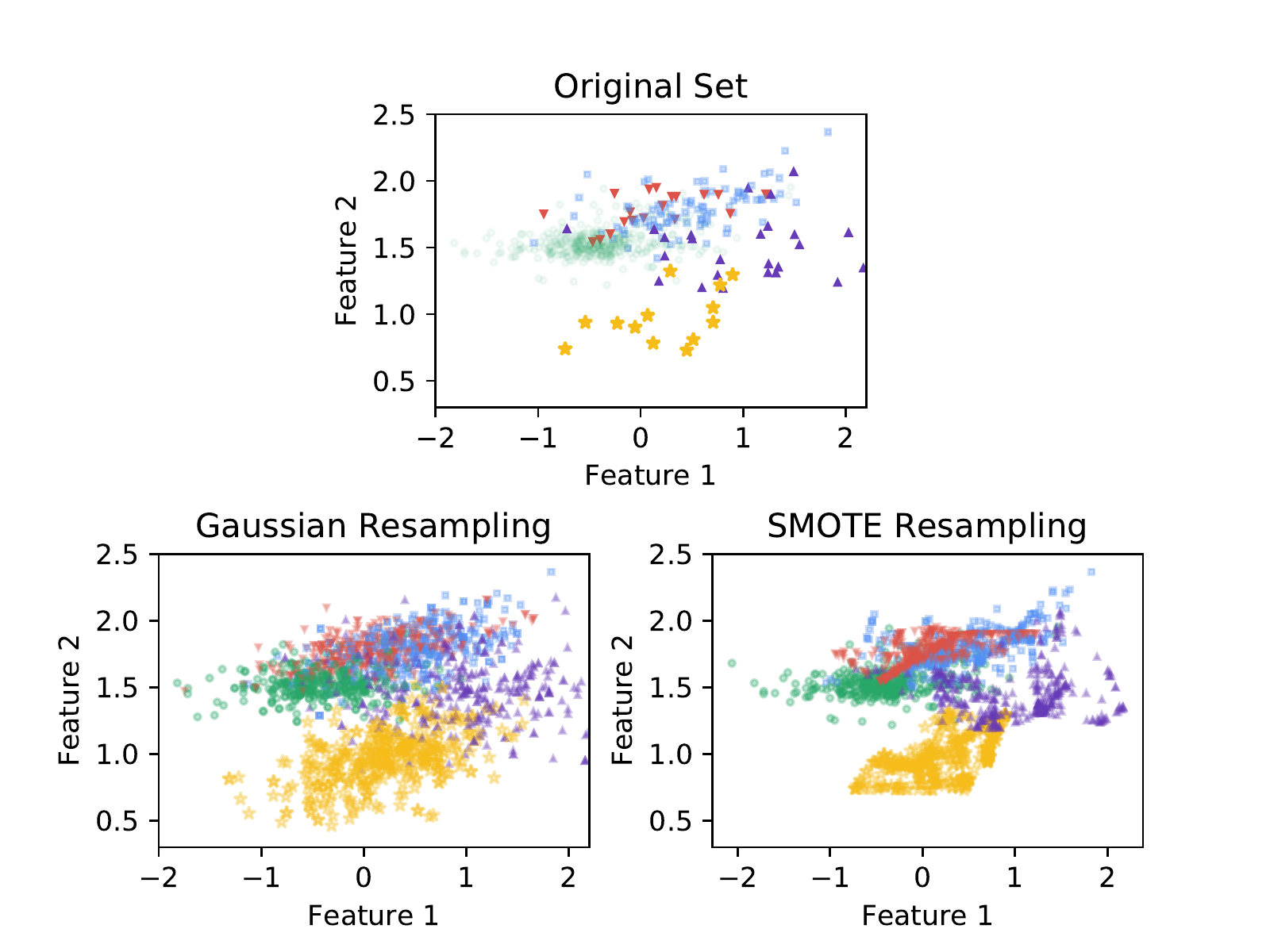}}
\caption{{\it Top:}  The original dataset plotted in terms of feature 1 versus feature 2, indicating both the span of the various SN classes and the imbalance in number of events per class. {\it Bottom:} The augmented dataset using MVG resampling ({\it Left}) and SMOTE resampling ({\it Right}).}
\end{center}
\label{fig:smote}
\end{figure*}

Here, we augment our training set with simulated events for two key reasons. First, our training set is unbalanced in terms of SN classes due to the differing observed rates of transients, with Type Ia SNe representing $\approx 70$\% of our sample (and more generally, of any magnitude-limited optical survey). Classification algorithms trained on unbalanced training sets tend to over-classify all objects as the dominant class. This is because the algorithms can minimize the decision-making complexity by ignoring minority classes in favor of correctly classifying the majority class. In our case, a classification algorithm may preferentially label all objects as Type Ia SNe to achieve an overall high accuracy.  Second, our training set is small in the context of machine learning, with the smallest class (SLSNe) containing just $17$ events.  

One approach to overcome this in the context of our method is to augment our training set with many draws from the MCMC posteriors. However, this would lead to clustering of solutions in feature-space that may bias the training algorithms. Instead, we address the issue of a small and imbalanced training set by synthesizing more event samples using two techniques.  First, we use the Synthetic Minority Over-sampling Technique (SMOTE; \citealt{chawla2002smote}) to over-sample all the non-Type Ia SN classes to be equally represented as the Type Ia SNe. SMOTE creates synthetic samples in feature space by randomly sampling along line segments joining the $k$ nearest neighbors of a sample, where $k$ is a free parameter of the algorithm. Here we find that $k=5$ performs well for sampling the minority classes. An example of the SMOTE resampling algorithm is shown in Figure~\ref{fig:smote}. A key feature of SMOTE resampling is that it produces realistic samples within each class, but it cannot produce samples outside the extent of the original sample. While this prevents the generation of unphysical models, it may overly constrain the properties of classes with only a few samples (e.g., SLSNe). 

Second, we augment the non-Type Ia SN classes by fitting the feature space of each class to a multivariate-Gaussian (MVG) and resampling from the fitted MVG. This is similar to the SMOTE algorithm in that it allows for the generation of new events that encompass a larger potential feature space. However, one key difference is that this method allows for synthesized events beyond the feature boundaries seen in the data. While this may lead to some unphysical models, it better reflects the potential spread in light curve parameters in poorly-sampled classes. An example of the MVG resampling is shown in Figure~\ref{fig:smote}.  

Both augmentation methods aim to increase our training set in a way which is representative of the set and therefore makes no attempt to correct for potential biases. This can potentially lead to increased misclassifications if our labelled training set is unrepresentative of a future test set; however, we expect no such effects within the training set of 513 objects.

\subsection{Classification}
\label{sec:classifiers}

Following the work of \citet{lochner2016photometric}, we test three classification algorithms: a support vector machine (SVM), a random forest (RF), and a multilayer perceptron (MLP). We optimize the hyperparameters of each algorithm independently using a grid search. Each algorithm and its tunable hyperparameters are described below. We use the {\tt scikit-learn} python package throughout the classification portion of our pipeline.

\subsubsection{Support Vector Machine (SVM)}
A SVM classifies the training set by finding the optimal hyperplane in feature space to minimize the number of misclassified samples. In particular, the SVM will select a hyperplane that maximizes the distance between class samples nearest the hyperplane (also known as the support vectors). In the majority of cases, the classes are not linearly separable within the feature space alone (i.e., there may be significant overlap between classes). Instead, the features are expanded into an infinite basis function using the so-called Kernel trick \citep{aizerman1964theoretical}, allowing one to find a feature space in which the separating hyperplane is linear. We optimize the kernel and a regularization term using a coarse grid search, allowing the kernel size to logarithmically range from $\sigma=1$ to $\sigma=100$ and the normalization to logarithmically range from 1 to 1000.  We find that a radial basis function kernel with width $\sigma=10$ typically results in optimal classification, with normalization values ranging depending on the pipeline.

\subsubsection{Random Forest (RF)}
RF classifiers \citep{breiman2001random} are built on the idea of a decision tree, which is a model that generates a set of rules to map input features to classes. This mapping is based on a series of branching decisions based on feature values (e.g., ``is the peak $g$-band magnitude brighter than $-19$?''). While single trees are theoretically sufficient for classification problems, they often lead to over-fitting due to specialized branching required for each class. Random forests overcome this problem by combining decision trees that are trained on different subsets of the training data and features. The ensemble of decision trees is then used as the classifier. There are a number of free parameters within a RF, including the number of decision trees, the number of nodes for each tree and the splitting rules for each node. Through a grid search of hyperparameters, we find that 100 decision trees utilizing the Gini impurity (the probability that a randomly chosen SN from a labelled class is misclassified) as a splitting criterion and allowing nodes to be split until all leaves are pure results in the highest accuracy.

\subsubsection{Multilayer Perceptron (MLP)}
A fully-connected MLP is the simplest artificial neural network (e.g., \citealt{schmidhuber2015deep}). It is composed of a series of layers of neurons, where each neuron is the dot product of the previous layer and a set of optimizable weights, passed through a nonlinear activation function.  A ``fully-connected'' MLP means that each neuron is connected to all neurons in the preceding layer. The nonlinear activation function is what allows a MLP to model nonlinear mappings between the feature set and classes. MLPs have many tunable parameters, including the number of layers, the number of neurons within each layer, the learning rate and a regularization term. We optimize the hyperparameters using a grid search, finding that two layers with ten neurons each typically performs best, and use the {\tt Adam} optimization algorithm \citep{adamopti2014} to train the MLP.

An example of a complete pipeline, excluding the MCMC fitting step, is available on GitHub\footnote{\url{https://github.com/villrv/ps1ml}}.

\begin{figure*}[t!]
\begin{center}
{\includegraphics[width=\textwidth]{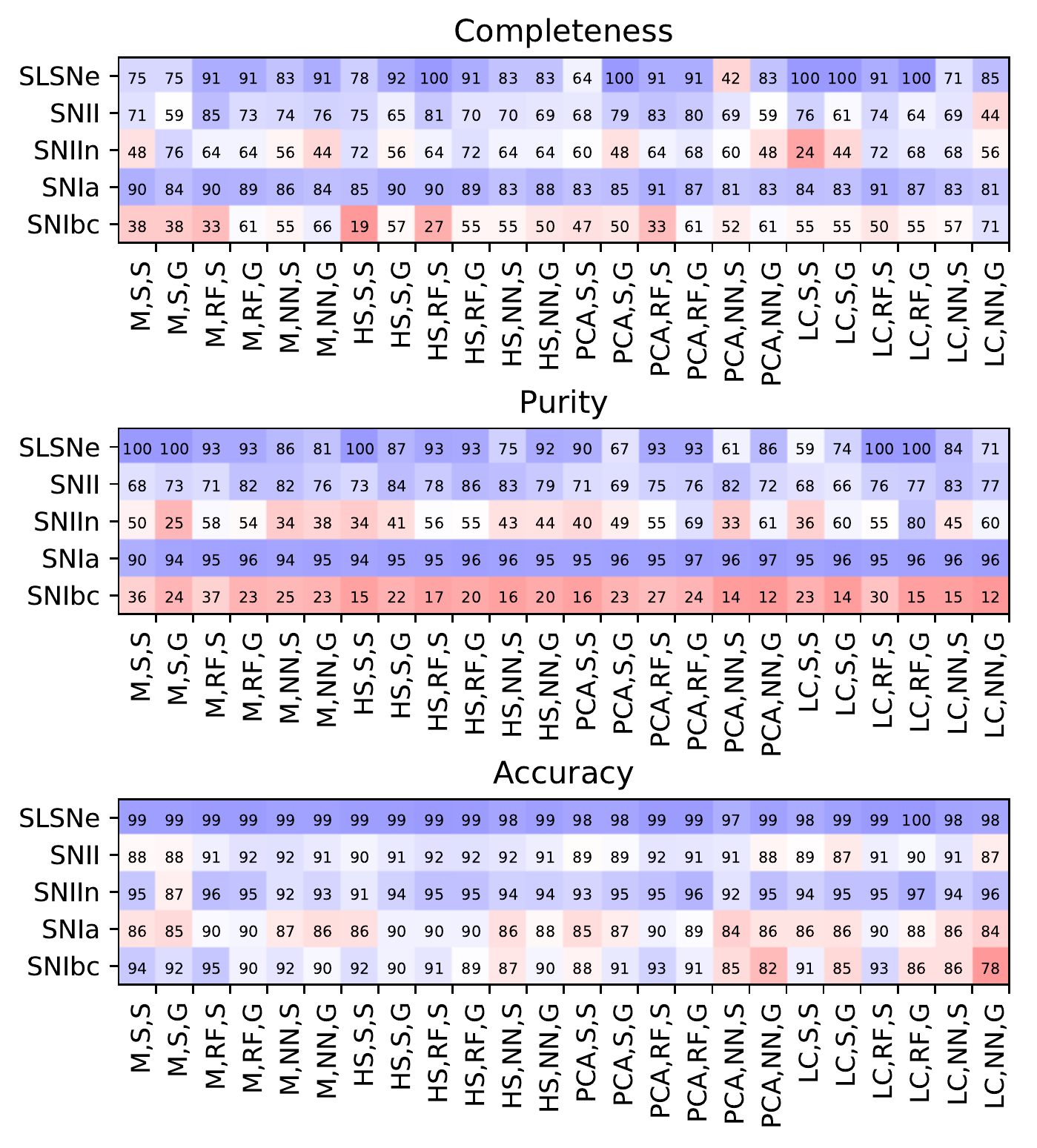}}
\caption{Completeness ({\it Top}), Purity ({\it Middle}) and Accuracy ({\it Bottom}) for each of the five spectrosopic SN classes across the 24 classification pipelines. Each pipeline is encoded by its feature extraction method (M, HS, PCA, LC), data augmentation method (S, G) and classification method (SVM, RF, NN).}
\end{center}
\label{fig:purcom}
\end{figure*}

\begin{figure*}[t!]
\begin{center}
{\includegraphics[width=\textwidth]{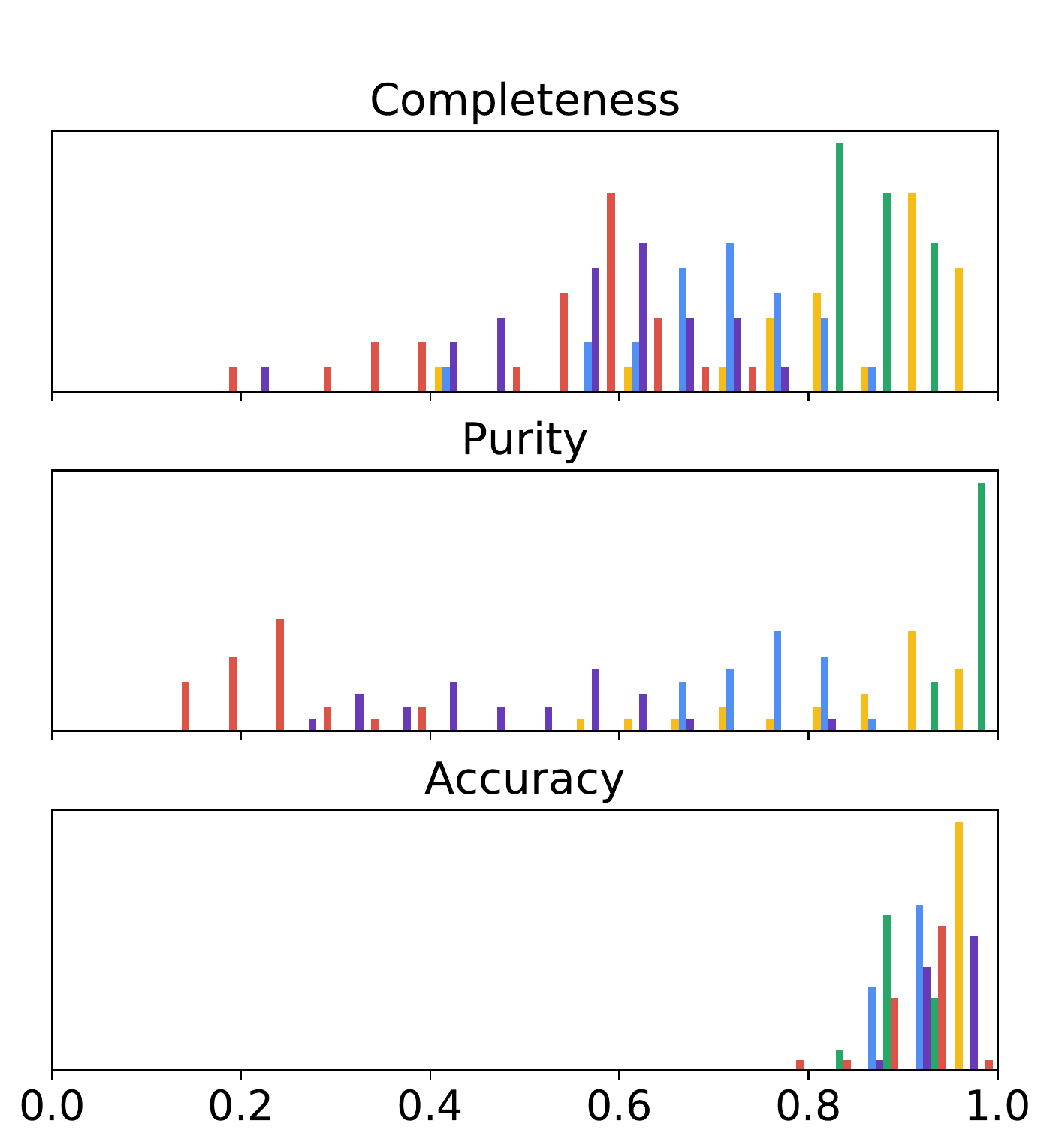}}
\caption{Histograms of Completeness ({\it Top}), Purity ({\it Middle}) and Accuracy ({\it Bottom}) for each of the five spectroscopic SN classes across the 24 classification pipelines.}
\end{center}
\label{fig:purcom_hist}
\end{figure*}

\begin{figure*}[t!]
\begin{center}
{\includegraphics[width=\textwidth]{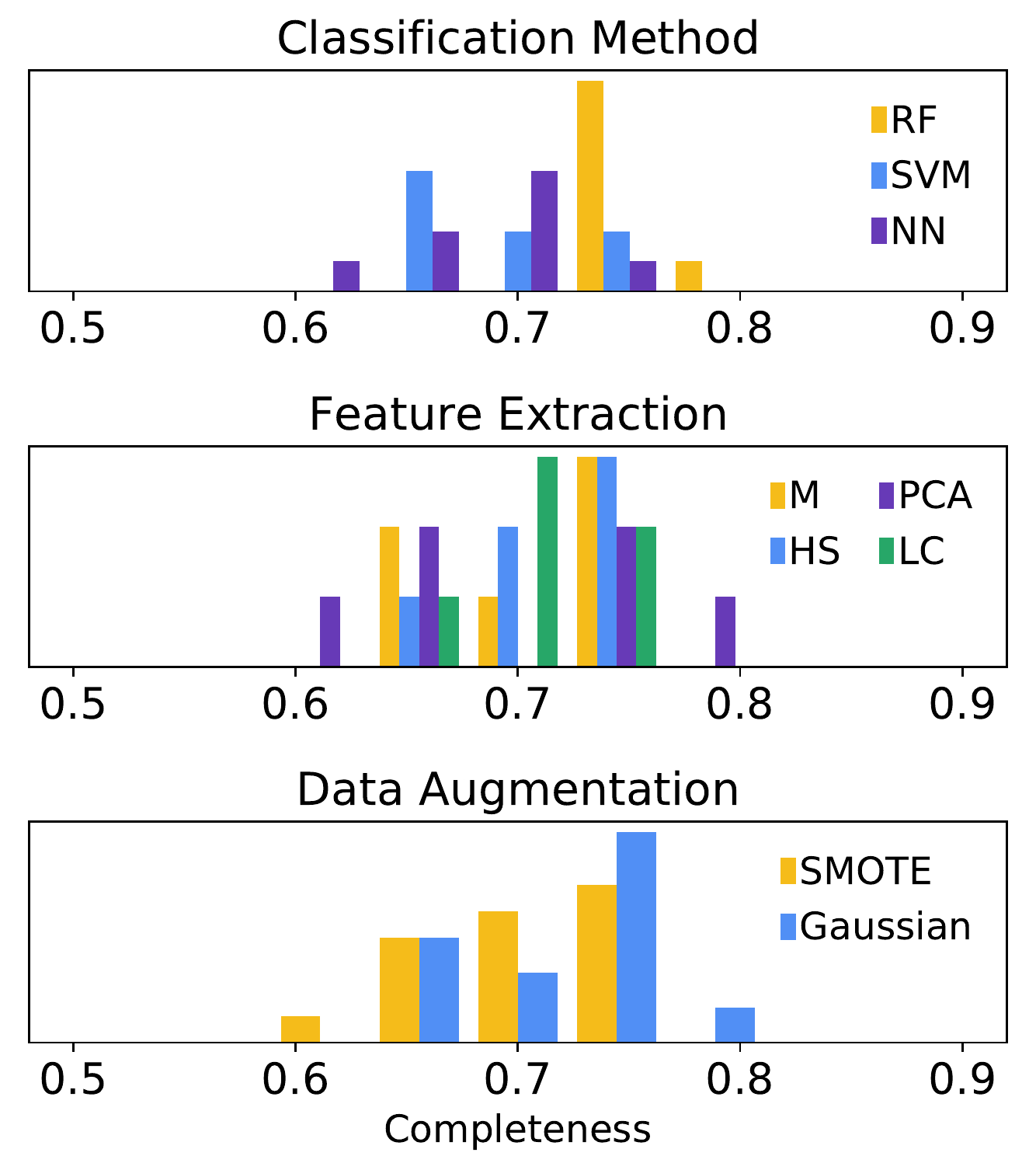}}
\caption{Histograms of completeness across all 5 SN classes, grouped by classification method ({\it Top}), feature extraction method ({\it Middle}) and data augmentation method ({\it Bottom}).}
\end{center}
\label{fig:hist_meth}
\end{figure*}

\section{Classification Results}
\label{sec:results}

\begin{figure}[t!]
\hspace*{-0.1in}
\vspace{-0.15in}
\begin{center}
{\includegraphics[width=0.45\textwidth]{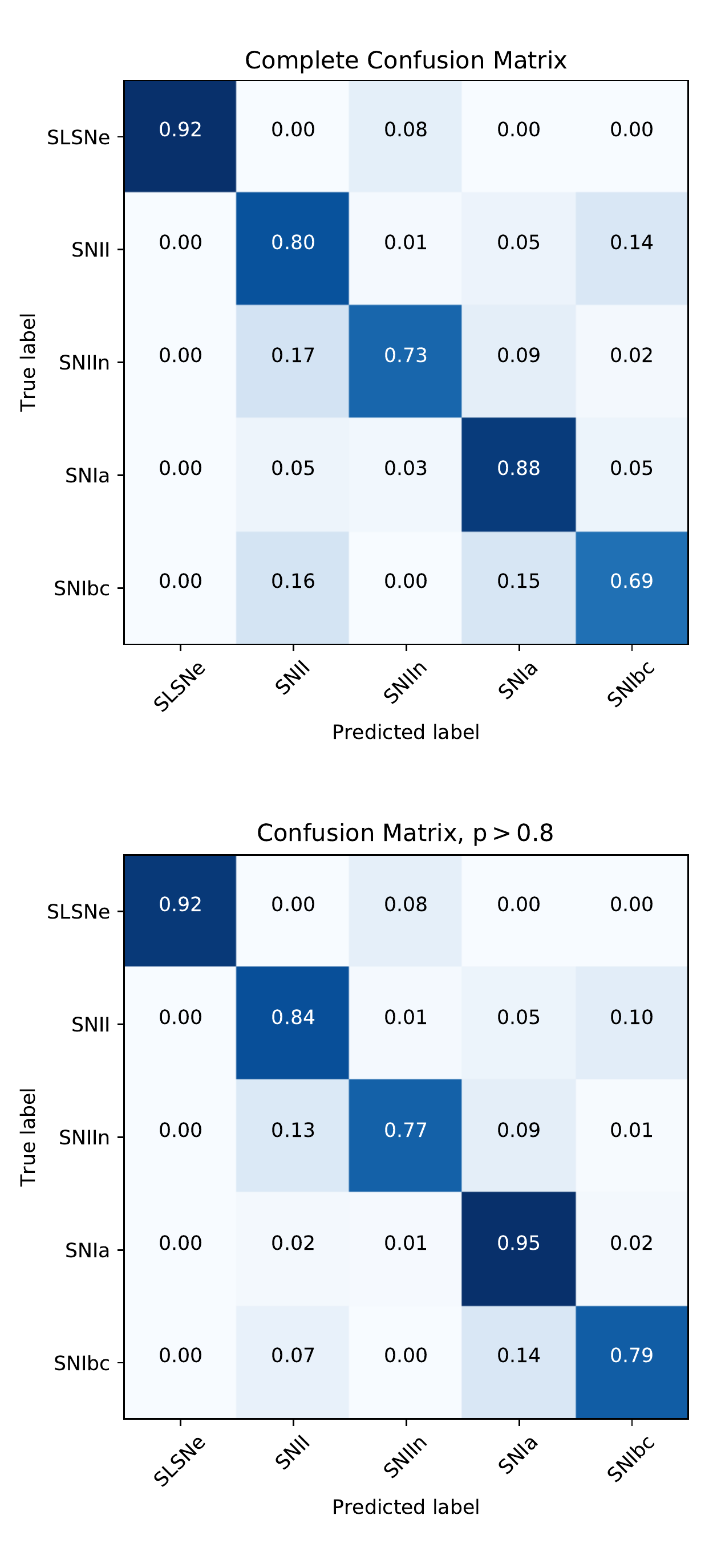}}
\vspace{-0.1in}
\caption{Confusion matrix for one of our best performing classification pipelines (PCA feature extraction, MVG data augmentation, and RF classifier) calculated using the full posterior distributions for each SN.  
We show the confusion matrix for both the full SN sample of 513 objects ({\it Top}) and only for the 429 events with a high classification confidence probability of $p>0.8$ ({\it Bottom}).}
\label{fig:confusion_matrix}
\end{center}
\end{figure}

\begin{figure*}[t!]
\begin{center}
{\includegraphics[width=0.95\textwidth]{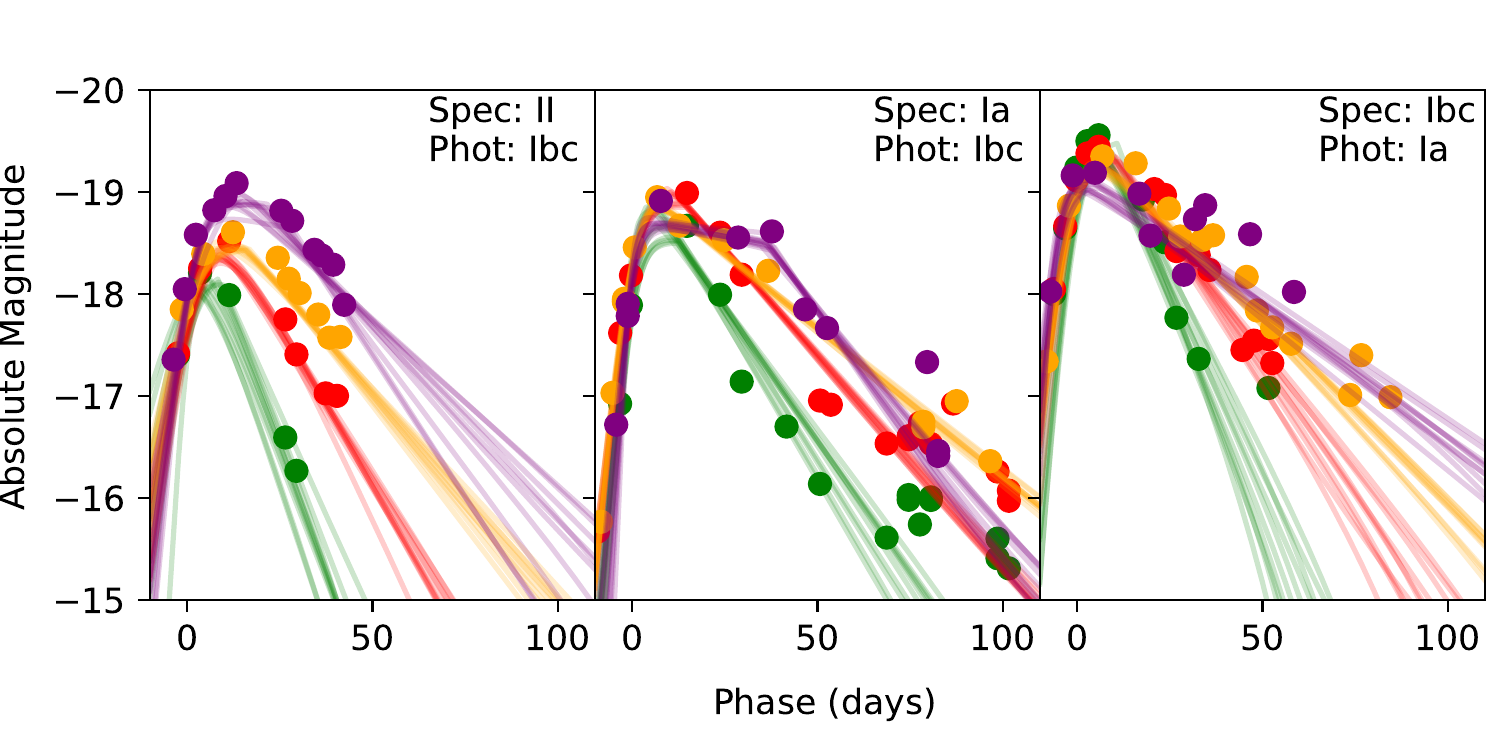}}
\caption{Light curves of three SNe classified incorrectly, but with high confidence ($p>0.9$). We note the spectroscopic and photometric classification of each event.  Given the high data quality, these misclassifications are due to inherent overlap between SNe in feature space.}
\label{fig:incor}
\end{center}
\end{figure*}

We combine each of the four feature extraction methods (M, HS, PCA, and LC), two data augmentation methods (SMOTE and MVG), and three classification algorithms (SVM, RF, and MLP) to test a total of 24 classification pipelines. For each pipeline, we use the full dataset to find the hyperparameters which optimize overall accuracy for the classification method. We optimize the hyperparameters over a coarse grid, due to the computational costs of performing a large grid search. We then perform leave-one-out cross-validation by iteratively removing one object from the sample, performing data augmentation on the remaining dataset, and training a classifier on the new set. We then test the trained classifier on the median posterior values of the removed object and record the predicted label. Due to computational costs, we only utilize the full posteriors for classification error estimation using our optimal pipeline.

\begin{figure*}[t!]
\begin{center}
{\includegraphics[width=\textwidth]{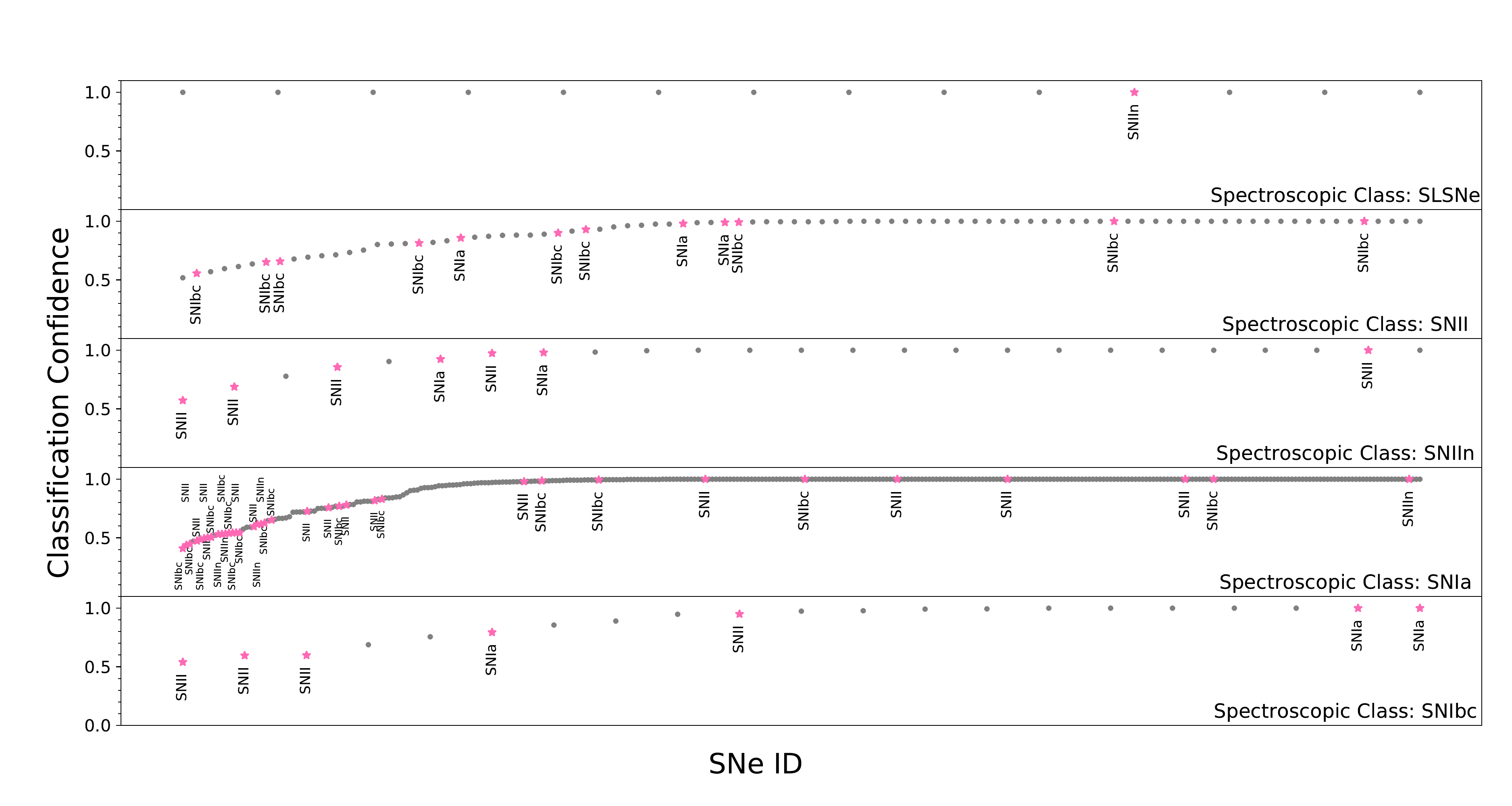}}
\vspace{-0.1in}
{\includegraphics[width=\textwidth]{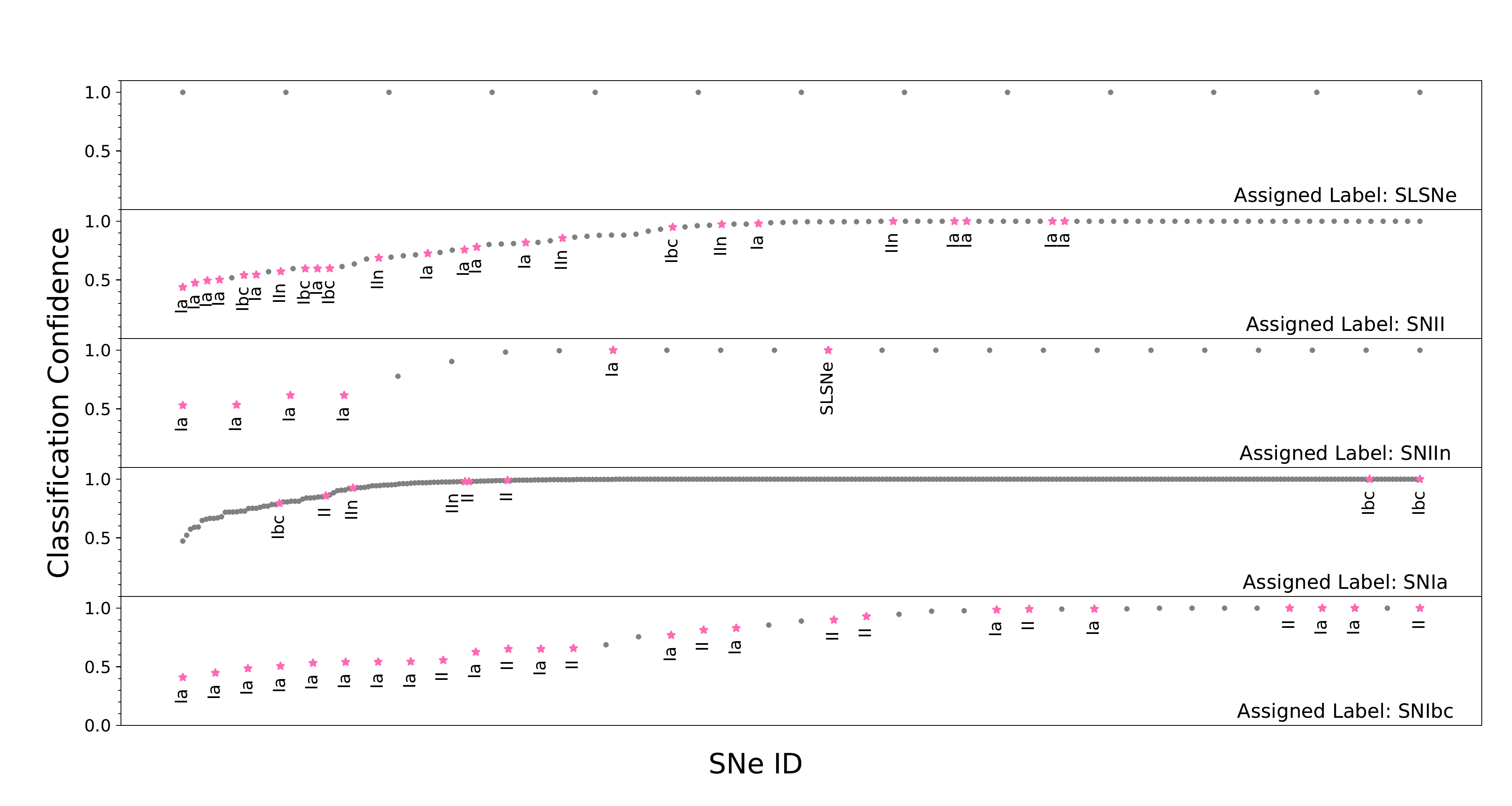}}
\caption{{\it Top:} For each {\it true} spectroscopic class, we show the correct classifications (grey) and misclassifications (pink), with the classification confidence plotted on the ordinate. The misclassified label is given next to each misclassified event.  We find that the bulk of the misclassifications are concentrated at low classification confidence. {\it Bottom:} For each \it{assigned} label, we again show the correct classifications (grey) and misclassifications (pink). The correct class label of each misclassified event is given. }
\label{fig:con_acc}
\end{center}
\end{figure*}

\subsection{General Trends}

In Figures~\ref{fig:purcom} and \ref{fig:purcom_hist} we plot the purity, completeness and accuracy for each of the 24 pipelines and each of the 5 SN classes. Figure~\ref{fig:purcom} provides a matrix representation with the percentage score noted for each combination of pipeline and SN class, while Figure~\ref{fig:purcom_hist} shows the same results in histogram format to aid in visualizing the range of completeness, purity, and accuracy values across the 24 pipelines for each SN class.  

We find that SLSNe and Type Ia SNe are consistently the classes with the highest purity and completeness, reaching $\gtrsim 90$\% for the best classification pipelines. This is due to the fact that SLSNe are easily separable from the other classes due to their high luminosity and longer durations (Figures~\ref{fig:redmag} and \ref{fig:dlps}), while Type Ia SNe are tightly clustered in feature space due to their intrinstic uniformity. 

In contrast, we find that Type Ibc SNe typically have the lowest purity and completeness, with $\approx 15-35\%$ and $\approx 25-65\%$, respectively, and a much wider spread in performance for the various pipelines.  The lower classification success rate is due to broader diversity within Type Ibc SNe, as well as their significant overlap with Type Ia SNe (e.g., Figure~\ref{fig:dlps}). 

For Type II SNe we find high values of purity and completeness of $\approx 65-85\%$ and $\approx 60-80\%$, respectively.  This overall high success rate is mainly due to the presence of a plateau phase that helps to distinguish most Type II SNe from the other classes.  However, the failed classifications are most likely due to the faster evolving Type II SNe (often called Type IIL), which tend to be misclassified as Type Ibc or Type Ia SNe due to overlap in light curve shapes (e.g., Figure~\ref{fig:dlps}).  

Finally, for Type IIn SNe we find purity and completeness of $\approx 30-80\%$ and $\approx 45-70\%$, respectively, reflecting the broad diversity of light curve morphologies and luminosities, with some events overlapping similar areas in feature space with Type Ia and Ibc SNe  (e.g., Figure~\ref{fig:dlps}).  As for the Type Ibc SNe, we find quite a broad dispersion in performance between the various pipelines.

For the overall accuracy across the 5 SN classes, we find generally high values of $\approx 100\%$ for SLSNe, $\approx 95\%$ for Type IIn SNe, $\approx 90\%$ for Type II SNe, $\approx 85-95\%$ for Type Ibc SNe, and $\approx 85-90\%$ for Type Ia SNe.  These values are essentially independent of the classification pipeline used.

To further explore the relative performance of the various pipelines, in Figure~\ref{fig:hist_meth} we plot the distribution of completeness across the full dataset, grouping the classification pipelines by feature extraction method, classification method, and data augmentation method.  We find that the classification method has the largest impact on completeness, with the RF classifiers performing noticeably better, and more uniformly, than the SVM and NN classifiers.  In terms of feature extraction we find that use of the model parameters (M) and PCA are somewhat advantageous compared to hand-selected (HS) features and the LC approach, although the PCA extraction leads to a broader range of outcomes.  Finally, the MVG augmentation method performs slightly better than SMOTE.  

The top three pipelines in terms of purity, completeness and accuracy share RF classification and PCA feature extraction, with both MVG and SMOTE augmentation. Between these pipelines, the overall accuracy differs by $\lesssim 5\%$ across the 5 SN classes.  In addition to performing well, the RF classifier also has the advantage of allowing us to measure the relative important of each feature.  For example, we test the relative importance of our hand-selected and model features in the RF classification pipeline using the ``gini importance'', a measure of the average gini impurity decrease across descending nodes \citep{leo1984classification}.  We find that the peak magnitudes are the most important interpretable features, with durations and other parameters being roughly equally important.

For simplicity, below we focus on the results of our pipeline with the highest purity ($72\%$) and completeness ($78\%$) scores with an average accuracy of $93\%$ across the 5 SN classes. This pipeline consists of PCA feature extraction, MVG data augmentation, and RF classifier; however, we emphasize that this pipeline does not significantly outperform the others. In Figure~\ref{fig:confusion_matrix} we present the final confusion matrix for this pipeline across the full training set. The confusion matrix is a quick-look visualization of how each class is correctly or incorrectly classified. We generate the confusion matrix using the full posteriors for each SN, so the probability densities have been effectively smoothed out across the matrix. To specifically assess the role of poor quality classifications, we show the confusion matrix for the full sample, as well as separately for classifications with a confidence of $p>0.8$ only (representing $\sim85$\% of the original sample).  In practice, one can optimize pipeline parameters to maximize sample purity, completeness or some other metric.

\subsection{Assessing Misclassifications}

Although the overall completeness for each SN class is high, we note several common misclassifications. First, Type II and Ia are the most likely classes to be misclassified as Type Ibc SNe.  The Type II SNe that are misclassified as Type Ibc SNe are typically either poorly sampled or are rapidly evolving (the so-called IIL events). Second, Type Ibc, IIn, and II SNe are the most likely classes to be misclassified as Type Ia SNe.  This is again due to specific events in those diverse classes that occupy the region in feature space that overlaps with the uniform Type Ia SNe.  Finally, Type IIn and Ibc SNe are the most likely classes to be misclassified as Type II SNe, again due to overlaps in feature space. Comparing the full sample to the subset of events with high classification confidence ($p>0.8$) we find that the fraction of misclassified events indeed declines (most notably for Type Ibc and Ia SNe), indicating that some misclassifications are simply due to poorly sampled light curves.  However, the overall trends for which classes are most likely to be misclassified as others remains the same, indicating that there is an inherent limitation to the classification success rate that is due to real overlaps in feature space.

We highlight several SNe that are misclassified, but with high confidence in Figure~\ref{fig:incor}. In these examples, a spectroscopic Type II SN with a rapid linear decline is misclassified as a Type Ibc SN; a slightly dim Type Ia SN is misclassified as a Type Ibc SN; and a fairly luminous Type Ibc SN is misclassified as a Type Ia SN. In each of these cases, the posterior of the fitted light curves is narrow, leading to little variability (i.e., a high confidence) in the final classification.  These events indicate that even with good photometric data quality there is inherent overlap of SNe in feature space that leads to misclassification.

The misclassifications of SNe are further highlighted in Figure~\ref{fig:con_acc}. Each panel in the top part of Figure~\ref{fig:con_acc} represents a spectroscopically classified class, while in the bottom part each panel represents a photometrically assigned class. The misclassified events in both cases are labeled to provide insight into the most common misclassification. In all panels the ordinate represents the overall classification certainty, based on many draws from the posteriors of each event. In all cases, the majority of misclassifications occur at the low confidence end ($p<0.8$), but there are also high confidence misclassifications. 

\begin{figure}[t!]
\begin{center}
\hspace*{-0.1in}
{\includegraphics[width=0.5\textwidth]{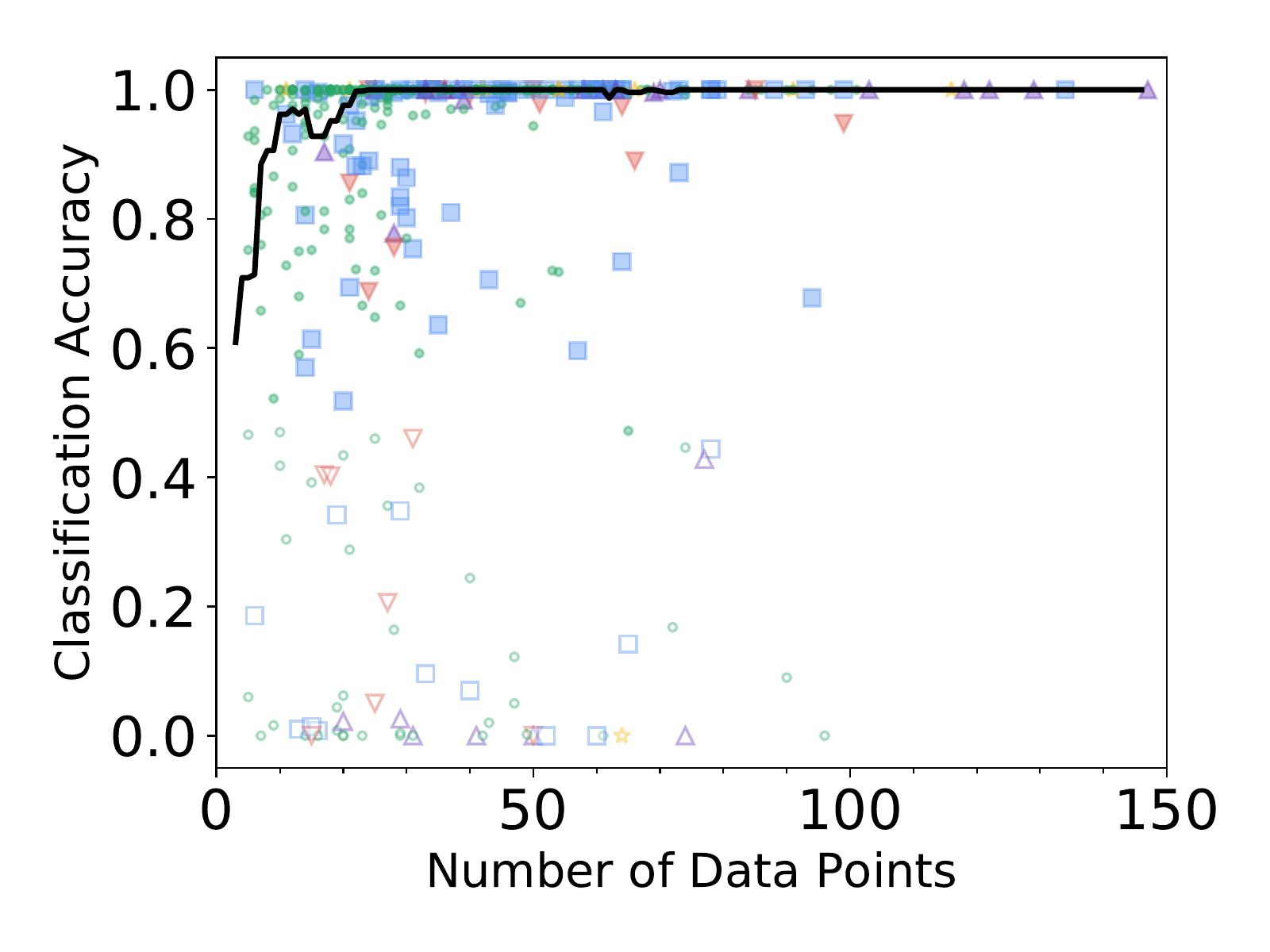}}
\caption{Classification accuracy as a function of number of light curve data points. The colors and shapes reflect the SN classes, and the black line represents a smoothed median to guide the eye. Filled symbols are SNe classified correctly, while open symbols are misclassified events.  We find that misclassifications are more prevalent for light curves with fewer points, but also that some events are misclassified even with tens of data points, as also highlighted in Figure~\ref{fig:incor}. }
\end{center}
\label{fig:info_content}
\end{figure}

We explore the role of data quantity in Figure~\ref{fig:info_content}, where we plot the classification accuracy as a function of total light curve data points for all 5 SN classes.  We again find that misclassifications are more likely in the regime of low number of data points, specifically $\lesssim 20$ data points.  However, as noted above, there are also high confidence misclassifications for events with a large number of data points.

\section{Comparison to Previous Photometric Classification Approaches}
\label{sec:comp}

The photometric classification of optical transients has been previously explored in the existing literature. Previous studies on machine learning methods have focused almost exclusively on the binary problem of Type Ia versus non-Type Ia SN classifications (e.g., \citealt{campbell2013cosmology,ishida2013kernel,jones2017measuring}), or have been trained and tested on simulated datasets (e.g., \citealt{kessler2010results,tonry2012pan,moller2016photometric,charnock2017deep,moller2019supernnova, muthukrishna2016deep}). We highlight the strengths and weaknesses of both approaches (which we note are disjoint) compared to our methodology. We emphasize that classification pipelines should ideally be compared using the same dataset and set of labels. No machine learning method, including the one presented in this paper, can be applied to a new test set without retraining or careful consideration of training-vs-test set biases. This is especially crucial when comparing our method to those created for simulated datasets, which have known biases, uncertainties and simulated physics.

Identification of Type Ia SNe from photometric light curves is essential for precision cosmology in the era of large photometric surveys \citep{scolnic2014systematic,jones2017measuring}, which is why many studies have specifically focused on Type Ia SN classification. However, the binary problem of Type Ia vs non-Typa Ia SN classification is much narrower (and simpler) than full classification of CCSN classes. As standardizable candles, Type Ia SNe are fairly homogeneous with observational variations (excluding reddening) that are well described by two observable features: stretch and peak luminosity. As a result, it is easier to separate the small area of feature-space corresponding to Type Ia SNe from other transients. Studies that focus on this approach achieve a classification accuracy of $\gtrsim 0.95$ (e.g., \citealt{ishida2013kernel,charnock2017deep,jones2017measuring,narayan2018machine,pasquet2019pelican}). Although our pipeline is trained and tested on an empirical dataset for 5 distinct SN classes, we find that our achieved purity ($\approx 95\%$), completeness ($\approx 90\%$) and accuracy ($\approx 95\%$) for Type Ia SN classification are actually comparable to methods that specifically train on the binary classification.  However, we note that studies such as \citet{moller2016} achieve this high purity rate without redshift information, which our method currently requires.

The vast majority of previous photometric classification studies used simulated datasets to train classifiers. This is largely due to the fact that few homogeneous photometric datasets with large numbers of spectroscopically-classified SNe exist. Most studies that train on simulated datasets use the Supernova Photometric Classification Challenge (SNPCC) training set \citep{kessler2010results}. The SNPCC dataset consists of 20,000 simulated SNe with $griz$ light curves, generated from templates of Type Ia, Ibc, IIP and IIn SNe (they do not include SLSNe). This dataset was presented as a community-wide classification challenge in preparation for the Dark Energy Survey, and was widely successful, with the top algorithms reaching an average Type Ia SN classification purity of $\approx 80\%$ and completeness of $\approx 95\%$. Works such as \citet{moller2019supernnova}and \citet{moss2018improved} have reported average classification accuracies of $\approx 90\%$ for CCSNe classes (similar to our reported accuracies here). Similarly, \citealt{lochner2016photometric} report an average Type Ia classification accuracy of $\sim84$\% using SALT2 light curve features. They further break down the CCSNe subclass into Type Ibc and Type II, where they report accuracies of $\sim63$\% and $\sim93$\%, respectively. We caution that the SNPCC dataset is not representative of the real diversity we encounter in on-going and future surveys, and should not be used as a benchmark for CCSN classification.  In particular, to generate synthetic light curves, \citet{kessler2010results} fit well-sampled real light curves from each CCSN class with a Bazin function.  Then they stretch Nugent CCSN templates\footnote{\url{https://c3.lbl.gov/nugent/nugent_templates.html}} to match the Bazin light curves. Variations within each class are included from both the sample of templates available and from random color variations derived from the Hubble scatter of Type Ia SNe and the peak luminosity derived from \citet{richardson2002comparative}.  While the collection of simulated Type Ia SNe likely samples the full phase-space of light curves, the non-Type Ia templates used to build the model light curve were severely limited. For example, only 2 Type IIn SN templates were used to generate 800 template light curves, and only 16 Type Ibc SN templates were used to generate 3,200 light curves. Because of this, we can expect methods that rely on this dataset to overestimate the accuracy of classifications for CCSN classes.  

A new classification challenge, PLAsTiCC \citep{allam2018photometric,plasticc}, is a more realistic simulated dataset that can be used as a benchmark for CCSN classification, although it too largely relies on theoretical models. Recent work by \citet{muthukrishna2019} find an average completeness of $\approx 65\%$ over the five SN classes that we have classified here (although we note that the PLAsTiCC challenge combines Type IIP/L and Type IIn SNe into one class). Our average completeness is significantly higher, at $\approx 77\%$.

\section{Limitations and Future Directions}
\label{sec:future}

The challenge of photometric classification for optical transients is broad and cannot be solved with one classification method alone. Like all methods, our classification pipeline aims to solve a simplified version of this problem: Given a complete light curve, a redshift, and a list of SN classes, what is the type of a given transient?  Here we highlight several improvements that can be made to our pipeline, and more broadly outline outstanding problems in the field of transient classification.

Our pipeline requires a redshift, which simplifies the problem of classification by anchoring the absolute magnitudes of every light curve. In our training set these redshifts were obtained from spectra of the transients and their host galaxies.  However, in the on-going and future surveys we expect that spectroscopic redshifts (from the SNe or host galaxies) will be rare.  On the other hand, LSST will provide photometric redshifts (photo-$z$) for all galaxies with $m<27.5$ mag, with an expected root-mean-square scatter of $\sigma_z/(1+z)\lesssim 0.05$ for galaxies with $m<25.3$ \citep{abell2009lsst}, and a fraction of outliers of $<10\%$ \citep{graham2017photometric}. A classification algorithm that can associate a transient to its host galaxy will therefore be able to utilize the photo-$z$ value. We anticipate that the additional uncertainty in the model fits due to the photo-$z$ uncertainty will not be a dominant factor. We additionally note that by including redshift information as a feature (even when doing so indirectly) we have limited the use of our pipeline to surveys of similar depth.

Additionally, our classification pipelines best utilizes full light curves, and are thus most naturally applicable for after-the-fact classification.  The most natural use is on the yearly samples of $\sim 10^6$ transients from LSST to enable large-scale population studies, as well as targeted studies of specific subsets (e.g., host galaxies of SLSNe).  While our method can work on partial light curves for real-time classification, its performance in this context is yet to be evaluated. Several studies that have explored the specific issue of real-time classification have found that recurrent neural networks perform well for this purpose (e.g., \citealt{charnock2017deep,moller2019supernnova,muthukrishna2019rapid}). 

Our algorithm currently relies exclusively on information derived from the transient light curves (other than the redshift).  However, useful contextual information about a SN can be extracted from the host galaxy. For example, SLSNe prefer low metallicity, dwarf galaxies \citep{lunnan2014hydrogen}, other CCSN classes span a wide range of star forming galaxies, and Type Ia SNe are found in both star forming and elliptical galaxies. Simple galaxy features, such as Hubble type, color, and SN offset can be easily incorporated into the classification pipeline (e.g.,\citealt{foley2013classifying}). This will be explored in follow-up work.

Furthermore, our algorithm is limited to classification within known SN classes (in this case 5 classes). To add additional classes under our current framework, we would need to incorporate new data into the training set and retrain the classification algorithms. Our pipeline is amenable to rapid training, so it is feasible to incorporate more classes in this way. For a more complex classification pipeline (e.g., one involving a large neural network), one could incorporate new classes cheaply using ``one-shot'' learning \citep{lv2006camera}, in which a classifier learns the characteristics of a new class using very limited examples. However, the addition of new classes will not solve the issue of how to identify unforeseen classes of transients and entirely new phenomena. Such a classifier is challenging to train, since outlier events are (by definition) rare.

Because our original training set is imbalanced and small, we needed to augment our dataset with simulated events drawn from the observed populations. For completeness, we test our best classification pipeline (PCA feature extraction and RF classifier) on the original training set without data augmentation. As expected, we find that we can classify classes with the most samples (Type Ia and II SNe) or those that are well-separated in feature space (SLSNe), as well as or better than our classification pipeline with data augmentation. However, the completeness of the minority classes, like Type Ibc and IIn SNe, falls by $20-40\%$. This is a good indication that data augmentation in the extracted feature space is a potential solution to the imbalanced classes. 

Our method neglects the possibility of a biased spectroscopic sample. For example, if the spectroscopic samples contains only the brighter end of the luminosity function for rare transients. In our presumed classification case in which we have access to the full light curves, one can use the full dataset to detect and minimize the effects of selection bias without knowing the true underlying distribution. For example, one can re-weight the importance of each SN in the spectroscopic training sample to better reflect the distribution of features from the full dataset (using, e.g., \citealt{huang2007correcting} and \citealt{cortes2008sample}). A detailed study of the effect of observational biases on transient classification is essential, but beyond the scope of this work.

Finally, we note that classification is only the first step in understanding the uncovered transients. Even for the currently rare SLSN class, LSST will discover $\sim 10^4$ events per year \citep{villar2018superluminous}. Additional data cuts that remove light curves with a minimal \textit{information content} (or those from which we cannot extract physical parameters) may be necessary in order to realistically fit a representative set of light curves.

\section{Conclusions}
\label{sec:conc}

Given increasingly large datasets and limited spectroscopic resources, photometric classification of SNe is a pressing problem within the wide scope of time-domain astrophysics.  Here we used the PS1-MDS spectroscopiccaly classified SNe dataset (513 events) to test a number of classification pipelines, varying the features extracted from each light curve, the augmentation method to bolster the training set, and the classification algorithms.  We used a flexible analytical model with an iterative MCMC process to model the g$_\mathrm{P1}$r$_\mathrm{P1}$i$_\mathrm{P1}$z$_\mathrm{P1}$ light curves of each event, and to generate posterior distributions.  We find that several pipelines (e.g., PCA feature extraction, MVG resampling, and RF classifier) perform well across the 5 relevant SN classes, achieving an average accuracy of about $90\%$ and a Type Ia SN purity of about $95\%$.

Our study is the first to use an empirical dataset to classify multiple classes of SNe, rather than just Type Ia versus non-Type Ia SN classification. Our overall results rival similar pipelines trained on simulated SN datasets, as well as those that utilize only a binary classification.  This indicates that we can utilize this approach to generate robust samples of both common and rare SN type (e.g., Type IIn, SLSNe) from LSST.

Finally, we highlight several areas for future exploration and improvement of our classification approach, including the use of contextual information and the possible application to real-time classification. We plan to extend this work and other classification approaches to the full set of PS1-MDS SN photometric light curves in future work.

\facilities{ADS, Pan-STARRs}
\software{Astropy \citep{astropy}, Matplotlib \citep{matplotlib}, NumPy \citep{numpy}, SciPy \citep{scipy}, Scikit-learn \citep{pedregosa2011scikit}, SNID \citep{Blondin2007}}

\acknowledgements
The Berger Time-Domain Group is supported in part by NSF grant AST-1714498 and NASA grant NNX15AE50G.  V.A.V.~acknowledges support by the National Science Foundation through a Graduate Research Fellowship. The UCSC team is supported in part by NASA grant NNG17PX03C; NSF grants AST--1518052 and AST--1815935; the Gordon \& Betty Moore Foundation; the Heising-Simons Foundation; and by a fellowship from the David and Lucile Packard Foundation to R.J.F. R.L. is supported by a Marie Sk\l{}odowska-Curie Individual Fellowship within the Horizon 2020 European Union (EU) Framework Programme for Research and Innovation (H2020-MSCA-IF-2017-794467). Some of the computations in this paper were run on the Odyssey cluster supported by the FAS Division of Science, Research Computing Group at Harvard University. The Pan-STARRS1 Surveys (PS1) and the PS1 public science archive have been made possible through contributions by the Institute for Astronomy, the University of Hawaii, the Pan-STARRS Project Office, the Max-Planck Society and its participating institutes, the Max Planck Institute for Astronomy, Heidelberg and the Max Planck Institute for Extraterrestrial Physics, Garching, The Johns Hopkins University, Durham University, the University of Edinburgh, the Queen's University Belfast, the Harvard-Smithsonian Center for Astrophysics, the Las Cumbres Observatory Global Telescope Network Incorporated, the National Central University of Taiwan, the Space Telescope Science Institute, the National Aeronautics and Space Administration under Grant No. NNX08AR22G issued through the Planetary Science Division of the NASA Science Mission Directorate, the National Science Foundation Grant No. AST-1238877, the University of Maryland, Eotvos Lorand University (ELTE), the Los Alamos National Laboratory, and the Gordon and Betty Moore Foundation.

\bibliography{mybib}

\end{document}